\documentclass[prb, twocolumn, showpacs, amsmath, amssymb]{revtex4-1} 
\usepackage{multirow}
\usepackage{bbm}
\usepackage{MnSymbol}
\usepackage{bbold}
\usepackage{booktabs}
\usepackage{epsfig}
\usepackage{graphicx}

\usepackage[dvips]{color}

\newcommand{\ix}[1]{\ensuremath{\text{#1}}} 
\newcommand{\K}{\ix{K}} 
\newcommand{\Ret}{\ix{ret}} 
\newcommand{\Adv}{\ix{adv}} 
\newcommand{\res}{\ix{res}} 
\newcommand{\hc}{\ensuremath{\text{H.c.}}} 
\newcommand{\abs}[1]{\ensuremath{\left| #1 \right|}} 
\newcommand{\Prin}[1]{\ensuremath{\mathcal{P} \left( #1 \right)}} 

\newcommand{\comm}[2]{\ensuremath{\left[ #1 , #2 \right]}} 
\DeclareMathOperator{\Tr}{Tr} 
\DeclareMathOperator{\Real}{Re} 

\begin{document}

\title{A renormalization group approach to time dependent transport\\ through correlated quantum dots}

\author{D.\ M.\ Kennes$^1$} 
\author{S.\ G.\ Jakobs$^1$}
\author{C.\ Karrasch$^2$} 
\author{V.\ Meden$^1$} 

\affiliation{$^1$Institut f\"ur Theorie der Statistischen Physik, RWTH Aachen
  University and JARA---Fundamentals of Future Information
  Technology, 52056 Aachen, Germany}

\affiliation{$^2$Department of Physics, University of California, Berkeley, California 95720, USA}

\date{\today}

\begin{abstract}

We introduce a real time version of the functional renormalization group which allows to study correlation effects on nonequilibrium transport through quantum dots. Our method is equally capable to address (i) the relaxation out of a nonequilibrium initial state into a (potentially) steady state driven by a bias voltage and (ii) the dynamics governed by an explicitly time-dependent Hamiltonian. All time regimes from transient to asymptotic can be tackled; the only approximation is the consistent truncation of the flow equations at a given order. As an application we investigate the relaxation dynamics of the interacting resonant level model which describes a fermionic quantum dot dominated by charge fluctuations. Moreover, we study decoherence and relaxation phenomena within the ohmic spin-boson model by mapping the latter to the interacting resonant level model.    

\end{abstract}







\maketitle


\section{Introduction}
\label{sec:introduction}

Obtaining a deeper understanding of nonequilibrium phenomena in the 
presence of many-body correlations is a major challenge 
in condensed matter physics. One particularly well-defined 
and controllable setup -- both from an experimental and 
a theoretical perspective -- are so-called quantum dots which feature a few correlated electronic degrees of freedom coupled to 
noninteracting leads. Their low-energy equilibrium physics is typically governed by the appearance of an energy 
scale which compared to the bare scales is strongly renormalized through the Coulomb interaction. A prominent example is the Kondo
effect:\cite{Hewson} If a (nearly) odd number of electrons reside on the dot, \textit{spin fluctuations} are strong below the Kondo temperature $T_K$. The latter depends exponentially on the bare system parameters given by the local 
charging energy $U$, the level-lead hybridization 
$\Gamma$ (`kinetic energy'), and the level position $\epsilon$. 
Another example are quantum dots dominated by correlated \textit{charge fluctuations}, in which the 
decay rate $\Gamma$ is renormalized. A prototypical model to describe such a scenario is the interacting resonant level model (IRLM).\cite{Schlottmann} 

In equilibrium, methods based on the renormalization group (RG) idea proved to be powerful tools to treat many-body quantum dot problems with a hierarchy of energy scales.\cite{poormansrg,rtrg,flowequations,nrg,Metzner11} Several RG(-like) approaches were therefore recently extended in order to address three different nonequilibrium scenarios. In the most simple case one is interested in the \textit{nonequilibrium steady state} induced by coupling to two (or more) leads that are held at chemical potentials differing by a bias voltage $ V $. The Hamiltonian itself is taken as time-independent. Prominent observables are the current $J$ through and the occupancy $\bar n$ of the dot levels. Problems of this class were studied using analytical RG(-like) methods such as the poor man's RG,\cite{Rosch} the real-time RG (RTRG),\cite{Herbert} and the flow-equation approach.\cite{flowequations} Numerical methods include Wilson's numerical RG (NRG) framework\cite{Anders1,Anders2} (note that it was recently questioned\cite{Roschneu} if NRG's inherent 
logarithmic discretization is reasonable in nonequilibrium), the time-dependent density-matrix renormalization group,\cite{PeterS,Fabian} an iterated path integral approach,\cite{Reinhold} and time-dependent quantum Monte Carlo\cite{Werner} (the last three are not RG-based).


A more intriguing (but certainly more complex) task is to study the \textit{relaxation dynamics} towards a steady state configuration, i.e.~to ask: How does a specific nonequilibrium state time-evolve under a still time-independent Hamiltonian? RG(-like) methods employed to address this question are the flow-equation approach,\cite{Kehrein} 
RTRG,\cite{Dirk,Herbert} and NRG.\cite{Anders1,Anders2} The last -- and for a theoretical 
description yet more challenging -- class of problems are those in which the Hamiltonian 
carries an \textit{explicit time dependence}; in the context of quantum dots, charge pumping is a typical example.\cite{pumpingwithoutcorrelations} The RTRG was recently extended to investigate this scenario.\cite{rtrgHt}    

The functional renormalization group (FRG)\cite{Metzner11} implements Wilson's RG idea in terms of an \textit{a priori} exact infinite hierarchy of differential flow equations for the many-body vertex functions. It has 
distinct general advantages over other Wilson-like RG procedures: Functional RG (i) can be applied directly to microscopic models and not only to effective field theories, (ii) provides information on all energy scales and not solely on the low-energy limit, and (iii) allows for a flexible introduction of the flow parameter (cutoff). The \textit{key approximation} is to truncate the infinite hierarchy at a given order. This is controlled for weak to intermediate interactions -- a parameter regime which for certain problems might still be dominated by electronic correlations. Indeed, the FRG was shown to provide a reliable tool to study the linear-response physics of single- and multi-level quantum dot setups.\cite{Karrasch06,Karrasch08,SeverinII,Karrasch10} It was recently extended\cite{SeverinI,Gezzi} in order to investigate the steady-state limit in nonequilibrium (i.e., to treat the first class of problems discussed above) where it captures certain aspects of nonequilibrium Kondo physics\cite{SeverinII} and yields a comprehensive picture of the finite-bias transport through a quantum dot dominated by correlated charge fluctuations.\cite{Karrasch10} The single impurity Anderson model as well as the IRLM were employed as prototypical examples.

This paper aims at a natural but nontrivial generalization of the functional RG framework which allows to tackle \textit{real time relaxation dynamics} as well as \textit{time-dependent Hamiltonians} of interacting quantum dot problems. In complete analogy with the prior extension of the FRG from linear response to steady-state nonequilibrium, this requires (i) to derive the exact hierarchy of flow equations from a functional that generates \textit{real-time} vertex functions, (ii) to introduce a cutoff which preserves symmetries (such as causality) in nonequilibrium, (iii) to consistently truncate the infinite hierarchy at a given order and to formulate a closed set of flow equation, and finally (iv) to implement an algorithm which solves them `numerically exact' in reasonable time. In contrast to all prior applications, devising such an algorithm is involved for the problem at hand. To this end, we organize our paper as follows:

In Sec.\ \ref{sec:Keldysh_formalism} we introduce a general Hamiltonian that describes an at this point unspecified quantum dot tunnel-coupled to noninteracting leads. We allow for time-dependent dot and tunnel parameters. In order to eventually set up the FRG flow equations, we discuss some basics of nonequilibrium single-particle Keldysh Green functions that depend on two time arguments. We shortly illustrate how to express the current $J$ and dot occupancy $\bar n$ in terms of those quantities.

In a next step (Sec.\ \ref{sec:FRG}) we derive the exact hierarchy of flow equations and discuss their form after truncating at the lowest nontrivial level. Even though this approximation -- which is the \textit{only} approximation within our approach -- can be strictly motivated for small Coulomb interactions only, it was successfully used to describe aspects of correlation physics for a variety of quantum dot setups in equilibrium\cite{Karrasch06} as well as in steady-state nonequilibrium.\cite{Karrasch10,SeverinII} We stress that the details of our FRG implementation for {\em fermions} are different from earlier extensions\cite{Gasenzer,PeterK} of the method to study the time evolution of interacting {\em bosons}. Importantly, we do not rely on the generalized Kadanoff-Baym ansatz to approximately solve the Dyson equation (which was used in Ref.\ \onlinecite{PeterK}) since we are sceptical about its general validity. 
 
In Sec.\ \ref{sec:irlm} we specify our (so far general) FRG approach for the interacting resonant level model. The latter describes a single spinless fermionic level with energy $\epsilon(t)$ that is locally coupled to two Fermi liquid leads via a tunnel matrix element $\tau(t)$ and a Coulomb interaction $U(t)$. In the steady-state limit, the corresponding flow equations could partially be solved analytically (in frequency space).\cite{Karrasch10a} This is no longer possible in our real-time representation, and we need to resort to a numerical treatment. Implementing a `numerically exact' solution efficiently on a standard computer is not straightforward. We present aspects of our algorithm in Sec.\ \ref{sec:num}.

Section\ \ref{sec:results} is devoted to elaborating the current $J(t)$ and the dot occupancy $\bar n(t)$ of the IRLM obtained from our FRG scheme. We focus on time-independent (but otherwise general) system parameters for reasons of simplicity but emphasize that the framework directly allows to study an explicit time dependence (results will be published elsewhere\cite{Dante}). In a nutshell, we find two different renormalized relaxation rates, characteristic oscillations with frequencies given by the level position 
relative to the left and right lead chemical potentials, and power-law corrections to the exponential time dependences. We do not observe the appearance of secular terms frequently encountered within perturbation theory.\cite{secularterms} A comparison of our results to real-time RG data\cite{Karrasch10a,Andergassen} strongly supports that the lowest-order FRG flow equations in real time capture hallmarks of correlation physics within the interacting resonant level model.

In absence of a bias voltage, the IRLM can be mapped onto the ohmic spin-boson 
model\cite{spinbosonrmp} whose relaxation dynamics we investigate as a second application (Sec.\ \ref{sec:spinboson}). 
We relate our results to predictions from field theory\cite{Saleur} and the so-called improved noninteracting blip approximation.\cite{Egger} 

A summary and a perspective for future applications of our time-dependent functional RG scheme are given in Sec.\ \ref{sec:conclusion}. In the Appendix we outline some technical details of the calculation of the Keldysh Green functions.


\section{Open Fermi systems in the Keldysh formalism}
\label{sec:Keldysh_formalism}

\emph{Hamiltonian} --- We aim at discussing time-dependent nonequilibrium transport through 
a quantum dot coupled to two or more leads (i.e., transport through an
open Fermi system). To this end, we introduce the Hamiltonian 
\begin{equation}
  H(t) = H^\ix{dot}(t) + \sum_\alpha [H^\res_\alpha + H^\ix{coup}_\alpha(t)] ~.
\label{fullH}
\end{equation}
The dot part constitutes of a single-particle term and a two-particle 
interaction,
\begin{align}
  \label{eq:H_dot}
  H^\ix{dot}(t) &= H^\ix{dot}_0(t) + H^\ix{int}(t),
  \\
  H^\ix{dot}_0(t) &= \sum_{ij} \epsilon_{ij}(t) d_i^\dagger d_j, \label{eq:H0}
  \\
  H^\ix{int}(t)  &= \frac{1}{4} \sum_{ijkl} \bar u_{ijkl}(t) d^\dagger_i
  d^\dagger_j d_l d_k,
\end{align}
where we employ standard second quantized notation. The reservoirs $\alpha$ are modelled as noninteracting,
\begin{equation}
  H^\res_\alpha = \sum_{k_\alpha} \epsilon_{k_\alpha}
  c^\dagger_{k_\alpha} c_{k_\alpha} , 
\end{equation}
and they are tunnel-coupled to the dot through
\begin{equation}
  H^\ix{coup}_\alpha(t) =  \sum_{k_\alpha,i}
  \gamma_{k_\alpha i}(t) c^\dagger_{k_\alpha}d_i + \hc.
\end{equation}
We explicitly allow for a time dependence of the parameters of $H^\ix{dot}$ and $H^\ix{coup}_\alpha$.

\emph{Initial statistics} --- We assume that the system is prepared using a product density matrix $\rho$ at time $t=0$ -- a situation which arises naturally when the dot and the reservoirs are decoupled for $t<0$. Furthermore, the reservoirs are supposed to initially be in grand canonical equilibrium with temperature $T_\alpha$ and chemical potential $\mu_\alpha$,
\begin{align}
  \rho(t=0) &= \rho_0 =  \rho^\ix{dot}_0 \otimes
  \rho^\res_{\alpha_1,0} \otimes \dots \otimes \rho^\res_{\alpha_m,0},
  \\
  \rho^\res_{\alpha,0} &= e^{-(H^\res_\alpha - \mu_\alpha
    N_\alpha)/T_\alpha} / \Tr e^{-(H^\res_\alpha - \mu_\alpha N_\alpha)/T_\alpha},
\end{align}
where $N_\alpha = \sum_{k_\alpha} c^\dagger_{k_\alpha} c_{k_\alpha}$. We choose units with $k_\ix{B} = 1$, $\hbar = 1$, and electron charge $e=1$. Finally, we assume that the statistical operator $\rho^\ix{dot}_0$ at $t=0$ allows for
the application of Wick's theorem\cite{danielewicz} and that its matrix representation commutes
with the initial single-particle dot Hamiltonian: $\comm{\rho^\ix{dot}_0}{\epsilon(t=0)} = 0$, where $\epsilon$ is the  matrix with entries 
$ \epsilon_{ij} $. For an initially empty quantum dot these 
requirements are trivially fulfilled; it is then irrelevant whether or not the two-particle 
interaction is present at $t=0$. However, when dealing with an initially nonempty quantum dot we have to assume that 
the interaction is turned on at time $t=0$ to avoid initial correlations. 

\emph{Green functions in Keldysh formalism} --- In order to describe the time evolution of the system for $t>0$, we employ the Keldysh formalism.\cite{HaugJauho,Rammer} All
single-particle properties of interest -- such as $J$ and $\bar n$ --  can be expressed in terms of the retarded 
and Keldysh component of the single-particle dot Green function,
\begin{align}
  \label{eq:GRet_def}
  G^\Ret_{ii'}(t,t') &= - i \Theta(t-t') \Tr \rho_0
  \left\{ d_i(t) , d_{i'}^\dagger(t') \right\} ,
  \\
  \label{eq:GK_def}
  G^\K_{ii'}(t,t') &= - i \Tr \rho_0
  \left[d_i(t) , d_{i'}^\dagger(t') \right].
\end{align}
The operators are in the Heisenberg picture with reference time
$t_0=0$; $\{ \ldots , \ldots \}$ refers to the anticommutator; 
$[ \ldots , \ldots ]$ denotes the commutator. Finally, the advanced Green function
\begin{equation}
  G^\Adv_{ii'}(t,t') =  i \Theta(t'-t) \Tr \rho_0
  \left\{ d_i(t) , d_{i'}^\dagger(t') \right\}  = G^\Ret_{i'i}(t',t)^\ast
\label{adjoint}
\end{equation}
is adjoint to the retarded one. All time arguments are positive throughout this paper, $t, t' > 0$.

It will prove useful to introduce Green functions that are computed w.r.t.~three different Hamiltonians: (i) the
noninteracting, decoupled dot propagator $g$ associated with $H^\textnormal{dot}_0$ only, (ii) the noninteracting but
reservoir dressed dot propagator $G^\ix{0}$ referring to $H$ with $ H^\ix{int}(t)=0 $, and (iii) the interacting and
reservoir dressed dot propagator $G$ calculated w.r.t.~the full $H$.

\emph{The noninteracting, decoupled dot propagator} can be obtained by computing $\partial_t g(t,t')$ and
$\partial_{t'} g(t,t')$ and then re-integrating with the correct boundary conditions:
\begin{align}
  \label{eq:gRet}
  g^\Ret(t,t') &= - i \Theta(t-t') {\mathcal T} e^{-i \int_{t'}^t \, dt_1 \, 
    \epsilon(t_1) },
  \\
  \label{eq:gK}
  g^\K(t,t') &= -i g^\Ret(t,0) (1-2 \bar n) g^\Adv(0,t').
\end{align}
${\mathcal T}$ denotes time ordering, and
\begin{equation}
  \bar n_{ii'} = \Tr \rho_0^\ix{dot} d^\dagger_{i'} d_i
\end{equation}
is the matrix of `occupancy' of dot states at time $t=0$.

\emph{The reservoir dressed but still noninteracting dot propagator} incorporates the presence of reservoirs via appropriate self-energy
contributions:  
\begin{align}
  \Sigma_\res &= \sum_\alpha \Sigma_\alpha ,
  \\
  \left[\Sigma^\ix{ret/K}_\alpha\right]_{i'i}(t',t) &=  \sum_{k_\alpha}
  \gamma_{k_\alpha i'}^\ast(t') g^\ix{ret/K}_{k_\alpha}(t',t) \gamma_{k_\alpha
    i}(t),
\end{align}
where the noninteracting reservoir propagator $g_{k_\alpha}$ is given in analogy to Eqs. \eqref{eq:gRet} and \eqref{eq:gK}. Since we are not interested in details of the reservoir band structure, we implement a continuous band of infinite width with a constant density of states (wide band limit),
\begin{equation}
  D_\alpha(\epsilon) = D_\alpha e^{-\delta \abs{\epsilon}},
\end{equation}
with $\delta \rightarrow 0^+$ assuring convergence of the energy integrals. 
This approach is widely used in the literature. Furthermore, we take the couplings between dot and
reservoirs to be independent of $k_\alpha$, that is $\gamma_{k_\alpha i}
= \gamma_{\alpha i}$. This gives
\begin{align}
  \label{eq:SigmaRet_wbl}
  \Sigma^\Ret_\alpha(t',t) &= - i \delta(t'-t) \Gamma_\alpha(t),
  \\
  \Sigma^\K_\alpha(t',t) &= - T_\alpha e^{- i \mu_\alpha(t'-t)}
  \Gamma_\alpha 
  \sum_\pm \frac{1}{\sinh[\pi T_\alpha(t'-t\pm i
    \delta)]},
  \label{eq:SigmaK_wbl}
\end{align}
where
\begin{equation}
  {\Gamma_\alpha}_{i'i}(t) = \pi D_\alpha \gamma_{\alpha i'}^\ast(t)
  \gamma_{\alpha i}(t).
\end{equation}
For the derivation of Eq.\ \eqref{eq:SigmaK_wbl} we exploited the
expansion
\begin{equation}
  1 - 2 \bar n_{k_\alpha} = - 2 T_\alpha \sum_{\omega_m} \frac{1}{i
    \omega_m - \epsilon_{k_\alpha} + \mu_\alpha},
\end{equation}
where the fermionic Matsubara frequencies $\omega_m$ are the odd multiples of
$\pi T_\alpha$, and the series is to be evaluated as a principal value
for $\abs{\omega_m} \rightarrow \infty$.  Note that the wide band
limit assumption of a continuous density of states implies the limit of
infinite reservoir size. Thus, this limit is to be performed before
further evaluating the dot propagator in order to properly define an open system configuration in which no recurrence phenomena occur. 

Dyson's equation for $G^0$ reads
\begin{equation}
  G^0 = g + G^0 \Sigma_\res g.
\end{equation}
The Green functions and the self-energy in this equation are matrices (retarded, advanced, and Keldysh 
components are ordered in the convention of Ref.\ \onlinecite{LarkinOvchinikov}),
\begin{equation}
  G^0 =
  \begin{pmatrix}
    G^{0, \Ret} & G^{0,\K}
    \\
    0 & G^{0, \Adv}
  \end{pmatrix}
  ,
  \quad
  \Sigma_\res =
  \begin{pmatrix}
    \Sigma_\res^\Ret & \Sigma_\res^\K
    \\
    0 & \Sigma_\res^\Adv
  \end{pmatrix}~,
\end{equation}
with each block being itself a matrix w.r.t.~the dot's single-particle quantum 
numbers. For the retarded component we find
\begin{equation}
  \label{eq:Dyson_Ret}
  G^{0,\Ret} = g^\Ret + G^{0,\Ret}
  \Sigma^\Ret_\res g^\Ret,
\end{equation}
where the multiplication abbreviates a summation over the dot quantum numbers as well as
integration over internal times,
\begin{equation}
  (AB)_{ii'}(t,t') = \sum_j \int_0^\infty d s A_{ij}(t,s) B_{ji'}(s,t').
\end{equation}
As $\Sigma^\Ret_\res \sim \delta(t'-t)$, the solution is simply
\begin{equation}
  \label{eq:G0Ret}
  G^{0,\Ret}(t,t') = - i \Theta(t-t') {\mathcal T} e^{- i \int_{t'}^t \, dt_1
    \left[\epsilon(t_1) - i \Gamma_\res(t_1) \right]}, 
\end{equation}
with $\Gamma_\res = \sum_\alpha \Gamma_\alpha$.  For the Keldysh component, Dyson's equation takes the form
\begin{multline}
  G^{0,\K} = g^\K + G^{0,\Ret} \Sigma^\Ret_\res
  g^\K + G^{0,\Ret} \Sigma^\K_\res g^\Adv
  \\
  + G^{0,\K} \Sigma^\Adv_\res g^\Adv,
\end{multline}
which is solved by
\begin{multline}
  \label{eq:G0K_contrib}
  G^{0,\K}(t,t') = - i G^{0,\Ret}(t,0) (1 - 2 \bar n)
  G^{0,\Adv}(0,t')
  \\
  +  (G^{0,\Ret} \Sigma^\K_\res
  G^{0,\Adv})(t,t'). 
\end{multline}

\emph{The full Green function} $G$ finally includes the
interaction. Its components satisfy
\begin{align}
  G^\Ret(t,t') &= G^{0, \Ret}(t,t') + \left[G^{\Ret} \Sigma^\Ret G^{0,
    \Ret}\right](t,t') \label{eins}
  \\
  G^\K(t,t') &= - i G^\Ret(t,0) (1 - 2 \bar n)
  G^\Adv(0,t') 
 \nonumber \\
  & \quad \qquad + [G^\Ret (\Sigma^\K_\res +
  \Sigma^\K) G^{\Adv}](t,t'),
  \label{eq:GK_contrib}
\end{align}
where $\Sigma$ is the self-energy associated with $H^{\rm int}$ (we will use the FRG to compute it approximately). The first
term in Eq.\ \eqref{eq:GK_contrib} governs the decay of the initial dot
occupancy while the second one describes how a new occupancy emerges under the influence of the reservoirs and the two-particle
interaction.

\emph{Physical observables} --- Single-particle properties of the system can be expressed in terms of the full Green function $G$. For instance, the (time-dependent) expectation value of the occupancy of a dot state easily follows from Eq.\ \eqref{eq:GK_def}:
\begin{equation}
  \bar n_i(t) = \frac{1}{2} - \frac{i}{2} G^\K_{ii}(t,t).
\label{occu}
\end{equation}
Another example is the current of particles leaving reservoir $\alpha$,
\begin{equation}
  J_\alpha(t) = - i \Tr \rho_0 \comm{H(t)}{N_\alpha(t)},
\end{equation}
where the operators are in the Heisenberg picture. A derivation similar to Ref.\ \onlinecite{meir} but in real time space yields
\begin{multline}
  J_\alpha(t) = - \Real \int_0^t d t' \Tr
  \big[\Sigma_\alpha^\Ret(t,t') G^\K(t',t)
  \\
  - G^\Ret(t,t') \Sigma^\K_\alpha(t',t)\big] .
\label{curr}
\end{multline}
Here, trace and multiplication abbreviate a summation over dot quantum numbers only.


\section{Functional renormalization group}
\label{sec:FRG}
The functional renormalization group is a quantum many-body method which allows to gain insights into the 
physics of interacting fermion and boson systems that exhibit a hierarchy of energy scales and/or 
competing instabilities.\cite{Metzner11} In a first step one supplements the noninteracting propagation by a flow parameter $\Lambda$, which for infrared divergent problems might reasonably be chosen as an infrared cutoff. Consequently, one-particle irreducible vertex functions (effective interactions) acquire a cutoff-dependence, and taking the derivative w.r.t.~the latter yields an exact infinite hierarchy of flow equations  -- in practice, this can be achieved using a generating functional. After truncating the hierarchy -- which can be done in a strictly controlled way -- one obtains a finite closed set of coupled differential equations for the self-energy (single-particle vertex), the effective two-particle interaction (two-particle vertex), and possibly higher order vertex functions (depending on the truncation order). The cutoff-free problem is recovered by integrating from $\Lambda=\infty$ where the vertices are known analytically down to $\Lambda=0$. Note that truncation is the only approximation; the full dependence on the single-particle quantum numbers and times (or frequencies) can be kept.     

Prior applications of the functional RG mainly focussed on the regime of linear response where one can conveniently resort to the Matsubara formalism (i.e., use imaginary times or frequencies). For quantum dot problems, two truncation schemes were employed. In the simplest approximation only the flow of the self-energy $\Sigma$ is taken into account; $\Sigma$ is then frequency independent.\cite{Karrasch06} Within this truncation all terms to 
{\it first order} in the two-particle interaction are kept, while higher-order terms are only partially included. We emphasize that the RG procedure enhances the quality of the approximation beyond 
perturbation theory (by including contributions from an infinite set of Feynman diagrams in a controlled way; see below). At the end of the RG flow the sum of the (frequency-independent) self-energy components merely correspond to an effective 
noninteracting problem,\cite{Karrasch06} which allows to gain intuitive insights into the observed correlation effects. This static approximation 
can be improved by incorporating the flow of the static part of the effective two-particle 
interaction; see Ref.\ \onlinecite{Karrasch06}. In a second truncation scheme -- which eventually includes all {\it second order} terms -- the full frequency and 
quantum number dependence of the two-particle vertex and the self-energy are 
kept, while the flow of the generated three-particle vertex is neglected.\cite{Karrasch08,SeverinII}  In this improved approximation a significantly larger number of coupled differential flow equations needs to be solved; this is only possible numerically and requires substantial computational effort. The second-order scheme was so far applied only to the single-impurity Anderson model.\cite{Karrasch08,SeverinII}   

The functional RG was recently extended to Keldysh frequency space in order to study the steady state of bias voltage driven quantum dots; both approximations mentioned above were employed.\cite{Gezzi,SeverinI,SeverinII,Karrasch10}      
The static lowest-order truncation is of limited usefulness if one aims at tackling dots that exhibit Kondo correlations.\cite{Gezzi} One needs to employ a higher-order 
scheme where the self-energy can acquire a frequency dependence.\cite{SeverinII} 
On the other hand, the static truncation allows to obtain a comprehensive picture of the nonequilibrium steady-state physics of the IRLM for small to 
intermediate interactions: Logarithmically divergent terms that show up in lowest-order perturbation theory are resummed consistently, and the RG-renormalized tunnel couplings features generic power laws with interaction-dependent exponents.\cite{Karrasch10} This in turn gives rise to highly nontrivial effects such as current-voltage characteristics dominated by power laws with interaction-dependent exponents. The latter can be computed to leading order in the interaction.\cite{Karrasch10a} The simplest truncation thus captures hallmarks of correlated charge fluctuations within the IRLM.  

\emph{Generating functional} --- In the following we sketch how to derive the FRG flow equations for 
time-dependent problems. The procedure is completely analogous to the one in equilibrium or steady-state nonequilibrium, and more technical details can be found in Refs.\ \onlinecite{DanteMaster}, \onlinecite{SeverinDiss}, and \onlinecite{ChristophDiss}. As a first step, we express the two-time Keldysh Green functions via their Keldysh contour functional integral 
representation:\cite{Kamenev}
\begin{widetext}
\begin{align}
\hat G^{pp'}_{ii'}(t,t')&=-i\int \mathcal{D}\bar\psi\psi \;\psi^{p}_{i}(t)\bar\psi^{p'}_{i'}(t')\, \exp\left\{i\int\limits_{0}^{\infty}d s \sum \limits_{i_1i_2}\sum \limits_{p_1p_2}\bar\psi^{p_1}_{i_1}(s+\eta)\left[\hat G^0 (s,s)^{-1}\right]_{i_1i_2}^{p_1p_2}
 \psi^{p_2}_{i_2}(s)-iS^{\text{int}} \right\},
\label{funint}
\end{align} 
\end{widetext}
where $ \psi $ denote Grassmann fields, and  $ p,p'=\pm $ are the usual Keldysh indices referring 
to the upper and lower branch of the Keldysh contour (they can be handled as additional 
quantum numbers for all practical purposes). The hat indicates the Green functions 
before rotation to the Keldysh basis (see Sec.\ \ref{sec:Keldysh_formalism}). The interacting part of the action is given by
\begin{equation}
S^{\text{int}}=\frac{1}{4}\sumint \limits_{\mathbbm{1},\mathbbm{2},\mathbbm{1'},\mathbbm{2'}} 
\bar u_{\mathbbm{1}\mathbbm{2}\mathbbm{1'}\mathbbm{2'}}\bar\psi_{\mathbbm{1}}\bar\psi_{\mathbbm{2}}\psi_{\mathbbm{2}'}\psi_{\mathbbm{1}'},
\end{equation}
using the multi-index $ \mathbbm{1}=(t,i,p) $ and the bare two-particle vertex
\begin{equation}
\begin{split}
\bar u_{\mathbbm{1}\mathbbm{2}\mathbbm{1}'\mathbbm{2}'}=&\delta(t_1-t_{1}')\delta(t_1-t_{2})\delta(t_1-t_{2}')\\
&\times \delta_{p_1,p_1'}\delta_{p_2,p_2'}\left(\sigma_z\right)_{p_1p_2} \bar u_{i_1 i_2 i_1' i_2'}(t) ,
\end{split}
\end{equation}
where $ \sigma_z $ denotes the $z$ Pauli matrix. The $m$-particle Green function
\begin{equation}
\begin{split}
\hat G&_{\mathbbm{1}\dots \mathbbm{m}\mathbbm{1'}\dots \mathbbm{m'}}=(-i)^m\\\times&\left \langle T_\gamma d^{p_1}_{i_1}[t_1]\dots d^{p_m}_{i_m}[t_m]d^{ \dagger\;p_{m'}}_{i_{m'}}[t_{m'}]\dots d^{\dagger\;p_{1'}}_{i_{1'}}[t_{1'}]\right\rangle_{\rho_{0}} ~.
\end{split}
\end{equation}
can be obtained from the following generating functional [the noninteracting part $S^0$ of the action is given by the first term in the exponential of 
Eq.\ (\ref{funint})]:
\begin{equation}
\mathcal{W}(\{\bar\eta\},\{\eta\})=\int \mathcal{D}\bar\psi\psi \; \exp\left\{S^0-iS^{\text{int}}-(\bar \psi ,\eta)-(\eta,\psi)\right\}
\end{equation}
through the derivative
\begin{equation}
\begin{split}
&\hat G_{\mathbbm{1}\dots \mathbbm{m}\mathbbm{1'}\dots \mathbbm{m'}}=\\&\left.(-i)^m\frac{\delta^m}{\delta \bar\eta_{\mathbbm{1}}\dots \delta \bar\eta_{\mathbbm{m}}}\frac{\delta^m}{\delta \eta_{\mathbbm{m'}}\dots \delta \eta_{\mathbbm{1'}}}\mathcal{W}(\{\bar\eta\},\{\eta\})\right|_{\eta=\bar\eta=0}.
\end{split}
\end{equation}
The corresponding functional that generates the one-particle irreducible vertex functions is given by the Legendre transformation
\begin{equation}
\Gamma (\{\bar\phi\},\{\phi\})=-\mathcal{W}^c (\{\bar\eta\},\{\eta\})-(\bar\phi,\eta)-(\bar \eta,\phi)+(\bar\phi,\left[ \hat G^0\right]^{-1} \phi)\label{GenVert1}
\end{equation}
of the generating functional of the connected Green functions:
\begin{equation}
\mathcal{W}^c(\{\bar\eta\}\{\eta\})=\ln\left[\mathcal{W}(\{\bar\eta\},\{\eta\})\right].
\end{equation}

\begin{figure}[t]
   \centering \includegraphics[width=0.65\linewidth,clip]{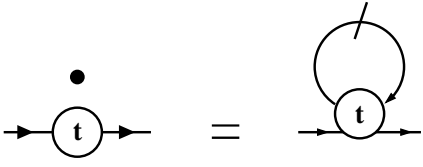}
\caption{Diagrammatic representation of the self-energy flow equation. The dot indicates the derivative w.r.t.~$ \Lambda $; the slanted line symbolizes the single-scale propagator. Within our first-order truncation scheme, the right-hand side is proportional to a delta function in time, and it is thus sufficient to introduce a single time argument $ t $.} 
\label{fig:Flowdia}
\end{figure}

\emph{Flow equations} --- If we supplement the free propagator by a (for the time being unspecified) flow parameter $\Lambda$, i.e., replace $ \hat G^0\to \hat G^{0,\Lambda}$, all 
vertex and Green functions acquire a $\Lambda$ dependence via Eq.\ (\ref{GenVert1}). Taking the derivative with respect to $ \Lambda $ yields the infinite hierarchy of FRG flow equations. Their general structure is to relate the $\Lambda$-derivative of the $m$-particle vertex to a certain set of diagrams involving the $m+1$-particle vertex and lower ones.\cite{Metzner11} This infinite hierarchy can generically only be solved by truncating it to a given order. Our approach uses the lowest-order scheme for reasons of simplicity; more elaborate approximations can in principle be devised straightforwardly. If the two-particle vertex is set to its bare value 
\begin{equation}
\gamma_2^{\Lambda}(\mathbbm 1,  \mathbbm 2, \mathbbm 1', \mathbbm 2')
=-i \bar u_{\mathbbm 1 \mathbbm 2\mathbbm 1' \mathbbm 2'}~,
\end{equation}
the only remaining flow equation is the one for the self-energy. It reads
\begin{eqnarray}
\partial_\Lambda\gamma_1^\Lambda(\mathbbm 1, \mathbbm 1') & = & 
\sum_{\mathbbm 2 , \mathbbm 2'}\left[\hat G^{\Lambda}\left(\partial_\Lambda 
[\hat G^{0,\Lambda}]^{-1}\right)\hat G^{\Lambda}\right]_{\mathbbm 2' \mathbbm 2}
\gamma_2^\Lambda(\mathbbm 1, \mathbbm 2,\mathbbm 1',\mathbbm 2')
 \nonumber \\ &= & - \sum_{\mathbbm 2 , \mathbbm 2'}\hat S_{\mathbbm 2' \mathbbm 2}^\Lambda
\gamma_2^\Lambda(\mathbbm 1 ,\mathbbm 2,\mathbbm 1', \mathbbm 2'),\label{eq:flow}
\end{eqnarray}
with the so-called `single-scale propagator' given by 
\begin{equation}
\begin{split}
\hat S^\Lambda_{\mathbbm 1 \mathbbm 1'} &= -\sum_{\mathbbm  2 , \mathbbm 2'}\hat G_{\mathbbm 1 \mathbbm 2'}^\Lambda\left[\partial_\Lambda [\hat G^{0,\Lambda}]^{-1}\right]_{\mathbbm 2' \mathbbm 2}\hat G_{\mathbbm 2 \mathbbm 1'}^\Lambda\\& = \partial^*_\Lambda \hat G_{\mathbbm 1  \mathbbm 1'}^\Lambda.
\end{split}
\end{equation}
We introduced the star differential operator $\partial^*_\Lambda$ which acts only on the free 
Green function $ \hat G^{0,\Lambda} $, not on $ \Sigma^\Lambda $, in the series expansion $ \hat G^{\Lambda}= \hat G^{0,\Lambda}+\hat G^{0,\Lambda}\Sigma^\Lambda \hat G^{0,\Lambda}+\dots$ . 
The flow equation is depicted diagrammatically in Fig.\ \ref{fig:Flowdia}. Even though its structure seems Hartree-Fock-like, it is important to stress that our approximation is in no way related to any mean-fieldish approach. The latter are known to suffer from severe artifacts in low-dimensional systems. 

\emph{Choice of a cutoff} --- It is one strength of the functional RG that it is not bound to a particular cutoff as long as the latter fulfills $ \hat G^{0,\Lambda=\infty}=0$ and $ \hat G^{0,\Lambda=0}= \hat G^0 $. A crucial step in the extension to nonequilibrium problems is to devise a scheme which conserves causality even after truncation.\cite{SeverinIII} This is guaranteed by the so-called hybridization flow; it was successfully used to study the steady state of the interacting resonant level model (see Refs.\ \onlinecite{SeverinII,Karrasch10} as well as the discussion above).             

The hybridization flow parameter $\Lambda$ can be physically interpreted as originating from coupling to additional auxiliary wide-band reservoirs (one for each dot state) with temperature $T_{\rm fp}$ and chemical potential $\mu_{\rm fp}$.\cite{SeverinII,Karrasch10} After rotating to the retarded, advanced and Keldysh basis, the corresponding self-energy entering the noninteracting but reservoir dressed dot propagator $G^0$ is determined by Eqs.\  \eqref{eq:SigmaRet_wbl} and \eqref{eq:SigmaK_wbl}:
\begin{align}
  \Sigma^\Ret_\ix{fp}(t',t) &= - i \delta(t'-t) \Lambda,
  \\
  \Sigma^\K_\ix{fp}(t',t) &= - T_\ix{fp} e^{- i \mu_\ix{fp}(t'-t)}
  \Lambda \sum_\pm \frac{1}{\sinh[\pi T_\ix{fp}(t'-t\pm i \delta)]},
\end{align}
with $\Lambda_{i'i} = \Lambda \delta_{i',i}$. We will frequently suppress 
the superscript $\Lambda$ (and only reinsert it when crucial to avoid misunderstanding). 
The flow starts at $\Lambda=\infty$ where the system instantaneously acquires its
stationary state ($\Lambda$ has the physical meaning of a decay rate). The initial conditions for the vertex
functions are therefore identical to those in the stationary state:\cite{SeverinII,Karrasch10}
$\gamma_n^{\Lambda=\infty}$ vanishes for $n \ge 3$,
$\gamma_2^{\Lambda=\infty}$ is given by the bare interaction vertex, and
\begin{align}
  \Sigma^{\ix{ret}, \Lambda=\infty}_{i'i}(t',t) &= \frac{1}{2}
  \delta(t-t') \sum_j \bar u_{i'jij},\label{eq:startvalueret}
  \\
  \Sigma^{\ix{K},\Lambda=\infty}_{i'i}(t',t) &= 0.
\end{align}
In prior applications of the hybridization cutoff to the IRLM, the temperature of the auxiliary leads was chosen equal to the physical one. In this work, we employ $T_{\rm fp}=\infty$ since (i) it implies $\Sigma^\K_\ix{fp} = 0$ which simplifies the flow equations, and -- more importantly -- (ii) it avoids any imprint of an artificial energy structure from the auxiliary leads and might thus retrospectively be a more reasonable choice on general grounds. Note that at infinite temperature the chemical potential $\mu_{\rm fp}$ no longer enters the flow equations. We have checked that steady-state results for the IRLM\cite{Karrasch10} are quantitatively unaltered if $T_\ix{fp} = \infty$ instead of $T_\ix{fp} = T$ is used. 

The Keldysh rotated single scale propagator [whose Keldysh component appears on the right-hand side 
of the flow-equation Eq.\ (\ref{eq:FlowRet})] is given by
\begin{equation}
  S = \partial_\Lambda^\ast G = (1 + G \Sigma) (\partial_\Lambda G^0)
  (1 + \Sigma  G).
\label{Sdef}
\end{equation}
Since $\partial_\Lambda G^0 = G^0 (\partial_\Lambda \Sigma_\ix{fp})
G^0$, we find
\begin{equation}
  S = G \frac{\partial \Sigma_\ix{fp}}{\partial \Lambda} G
\end{equation}
with components
\begin{equation}
  \label{eq:SRet}
  S^\Ret = G^\Ret \frac{\partial
    \Sigma^\Ret_\ix{fp}}{\partial \Lambda} G^\Ret
  = - i  G^\Ret G^\Ret
\end{equation}
as well as
\begin{align}
  S^\K &= G^\Ret \frac{\partial
    \Sigma^\Ret_\ix{fp}}{\partial \Lambda} G^\K + G^\Ret \frac{\partial
    \Sigma^\K_\ix{fp}}{\partial \Lambda} G^\Adv + G^\K \frac{\partial
    \Sigma^\Adv_\ix{fp}}{\partial \Lambda} G^\Adv
  \nonumber \\
  & = - i  G^\Ret G^\K + i  G^\K G^\Adv +  G^\Ret \frac{\partial
    \Sigma^\K_\ix{fp}}{\partial \Lambda} G^\Adv.
  \label{eq:SK}
\end{align}
Finally, the flow equation \eqref{eq:flow} translates to
\begin{align}
&\partial_\Lambda \Sigma^{\text{K,}\Lambda}=0 , \\
&\partial_\Lambda \Sigma^{\text{ret},\Lambda}_{i_1 i_{1'}}(t',t)=\partial_\Lambda \Sigma^{\text{adv},\Lambda}_{i_1 i_{1'}}(t',t)\notag\\&=-\sum\limits_{i_2,i_{2}'}S^{\text{K},\Lambda}_{i_2' i_2}(t,t)\left(-i \bar u_{i_1 i_2 i_1' i_2'}(t)\right)\delta(t'-t).\label{eq:FlowRet}
\end{align}
Importantly, the self-energy component $ \Sigma^\K $ does not flow.

\emph{Explicit form of the flow equation} --- In the lowest order truncation scheme, the expressions for $S$ can be simplified further:
\begin{align}
  \Sigma^\Ret(t',t) &= \delta(t'-t) \Sigma^\Ret(t),
  \\
  \Sigma^\K (t',t) &= 0,
\end{align}
and hence
\begin{equation}
\label{eq:GRtdependent}
  G^{\Ret}(t,t') = - i \Theta(t-t') {\mathcal T} e^{- i \int_{t'}^t \, d t_1 
    \left[\epsilon(t_1) - i \Gamma_\res(t_1) - i \Lambda +
      \Sigma(t_1) \right] },
\end{equation}
which in turn entails the multiplication formula
\begin{equation}
\label{eq:multiformula}
  G^{\Ret}(t,t') G^{\Ret}(t',t'') = - i \Theta(t-t')
  \Theta(t'-t'') G^{\Ret}(t,t'').
\end{equation}
Applying these properties to Eqs.\ \eqref{eq:SRet} and \eqref{eq:SK} eventually yields
\begin{equation}
  S^\Ret(t,t') = (t'-t) G^\Ret(t,t')
\end{equation}
as well as
\begin{multline}
  \label{eq:SK_trunc}
  S^\K(t,t') = - (t+t') G^\K(t,t')
  \\
  + \int_0^\infty  \, dt_1\,t_1 \Big\{G^\Ret(t,t_1) \big[(\Sigma^\K_\res +
    \Sigma^\K_\ix{fp}) G^\Adv \big](t_1,t')
    \\
    + \big[G^\Ret(\Sigma^\K_\res +
    \Sigma^\K_\ix{fp})\big](t,t_1) G^\Adv(t_1,t') \Big\} 
  \\
  + \left[ G^\Ret \frac{\partial \Sigma^\K_\ix{fp}}{\partial \Lambda} G^\Adv \right](t,t'),
\end{multline}
where the last term vanishes for our choice $ T_{\ix{fp}}=\infty $.

We conclude with two comments: (i) The right-hand side of the flow equation (\ref{eq:FlowRet}) depends on the Green function via Eq.\ (\ref{eq:SK_trunc}); thus, the retarded self-energy is fed back into the flow through the Dyson equation(s) (\ref{eins}) and (\ref{eq:GK_contrib}). This illustrates that Eq.\ (\ref{eq:FlowRet}) actually is a differential equation whose integration requires the solution of Eq.\ (\ref{eq:SK_trunc}) and Dyson's equation for each $\Lambda$. (ii) The self-energy originating from the two-particle interaction appears in combination with the single-particle part of the dot Hamiltonian in all equations at hand. Thus, the former can be intuitively interpreted as a time-dependent renormalization of the effective single-particle parameters. Note that this immediately illustrates that the continutity equation, which for the most relevant case of two leads $\alpha=L,R$ reads
\begin{equation}
J_L(t)+J_R(t)=\frac{d}{dt}\sum\limits_i \bar n_i(t),
\end{equation}
holds within our approximation. 

\emph{Brief summary} --- We have introduced a real-time FRG formalism that is capable to describe the time evolution of interacting quantum dots 
coupled to Fermi liquid reservoirs on all scales from transient to asymptotic. The dot parameters as well as the level-lead hybridizations can carry an explicit time dependence. This equally allows to study the transient dynamics of a bias voltage driven setup under a time-independent Hamiltonian 
as well as problems that exhibit time-dependent parameter variations (e.g., charge pumping). Our derivation of the FRG flow equations was kept general: We did not specify the number of correlated dot levels and leads or the actual geometry determined by the level-lead couplings. As an application we will now investigate the relaxation dynamics of the IRLM in the presence of a finite bias voltage. Results for explicitly time-dependent Hamiltonians will be presented elsewhere.\cite{Dante}


\section{Interacting resonant level model}
\label{sec:irlm}
\begin{figure}
  \centering \includegraphics[width=0.8\linewidth,clip]{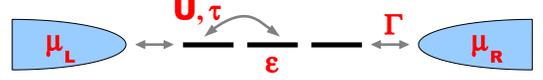}
  \caption{(Color online) Sketch of the three-site version of the interacting resonant level model.}
  \label{fig:irlm}
\end{figure}

The dot Hamiltonian of the frequently-mentioned interacting resonant level model (see Fig.~\ref{fig:irlm} for a sketch) is given by 
\begin{align}
  H^\ix{dot}_0(t) &= \epsilon n_2 - U\left(\frac{n_1}{2} + n_2 +
    \frac{n_3}{2}\right) ,  \nonumber
  \\
  & \hspace{6em} + \tau(d_1^\dagger d_2 + d_2^\dagger d_3 + \hc)\label{eq:singlepartinter}
  \\
  H^\ix{int}(t) &= U (n_1 n_2 + n_2 n_3),\label{eq:twopartinter}
\end{align}
where $n_i = d_i^\dagger d_i$ denote three spinless fermionic levels connected locally through a hopping matrix element $\tau >0$ and a Coulomb interaction $U$ (the latter will mainly be taken as repulsive; for an exception see Sec.\ \ref{sec:spinboson}). Only the central site 2 can be moved in energy by changing the `gate voltage' $\epsilon$. The second term in the single-particle part of the Hamiltonian is added for mere convenience so that $ \epsilon=0 $ corresponds to the point of particle-hole symmetry. This term is incorporated into the self-energy; it cancels the initial condition Eq.\ \eqref{eq:startvalueret} of the self-energy flow.
  
Dot sites 1 and 3 are coupled to left ($\alpha=L$) and right ($\alpha=R$) noninteracting wide-band leads. Their chemical potentials differ by an applied bias voltage $V = \mu_\ix{L} - \mu_\ix{R} \geq 0$ which we chose symmetrically as $ \mu_\ix{L}=-\mu_\ix{R} = V/2$ for reasons of simplicity. We focus exclusively on the zero-temperature limit. The leads give rise to a self-energy term $\Sigma_\res = \Sigma_\ix{L} + \Sigma_\ix{R}$ that reads [see Eqs.\ \eqref{eq:SigmaRet_wbl} and \eqref{eq:SigmaK_wbl}]
\begin{align}
  \left[ \Sigma^\Ret_\res \right]_{\, ii'}(t',t) &= - i \delta(t'-t)
  \delta_{i,i'}(\delta_{i,1}+\delta_{i,3}) \Gamma,
  \\
  \left[ \Sigma^\K_\res \right]_{\, ii'}(t',t) &= - \frac{2}{\pi}
  \Prin{\frac{1}{t'-t}} \delta_{i,i'} \nonumber
  \\
  & \quad \qquad \times \left(\delta_{i,1} e^{-i \mu_\ix{L}(t'-t)} +
    \delta_{i,3} e^{-i \mu_\ix{R}(t'-t)} \right) \Gamma,
\label{eq:SigmaKinter}
\end{align}
where $\mathcal{P} $ denotes the principal value, and
\begin{equation}
  \Gamma = {\Gamma_\ix{L}}_{11} = {\Gamma_\ix{R}}_{33}.
\end{equation}
We have chosen (again for reasons of simplicity) spatially-symmetric hoppings and interactions but emphasize that these restrictions can be abandoned at the expense of minor additional effort.

In the literature one frequently encounters a field-theoretical realization of the IRLM (see Refs.\ \onlinecite{Schlottmann,Karrasch10a,Doyon,Andergassen,Schmitteckert} as well as references therein). The latter consists of a single spinless fermionic level which is coupled 
to two leads. A two-particle interaction acts between the fermion occupying the level and the ones 
located at the boundaries of the leads. Our microscopic three site dot model is equivalent to the field-theoretical IRLM in 
the so-called scaling limit  
\begin{equation}
\tau,|\epsilon|,|U|,V \ll \Gamma  \label{eq:Scaling limit}
\end{equation}
where the first and third site of the dot can effectively be incorporated into the reservoirs as $\Gamma$ is much larger than all other (bare) energy scales (see Sec.\ \ref{sec:resultsnoint} for more details). Within our functional RG approach we cannot directly treat a single site model with density-density interaction to the leads.
 
\emph{The equilibrium physics} of the IRLM is dominated by an interaction-dependent renormalization of the hopping $\tau$. First order perturbation theory leads to a logarithmic term in the Matsubara self-energy $\Sigma^{\rm M}$ which for $\epsilon=0$ (half filling) reads\cite{Karrasch10} 
\begin{eqnarray}
\frac{\Sigma^{\rm M}_{1,2}}{\tau}  = 
\frac{\Sigma^{\rm M}_{2,3}}{\tau}  = - \frac{U}{ \pi \Gamma} 
\ln{\left( \frac{2 \tau^2}{\Gamma^2} \right)} \; .
\end{eqnarray}   
This logarithm is (automatically) resummed by equilibrium functional RG which analytically yields\cite{Karrasch10,Doyon}  
\begin{align}
\frac{\tau^{\rm ren}}{\tau} \sim \left\{\begin{array}{ccl}
\left(\frac{\tau}{\Gamma}\right)^{-2 U /(\pi \Gamma)+{\mathcal O}(U^2)}&& |\epsilon| \ll \tau \ll \Gamma\\
\left( \frac{|\epsilon|}{\Gamma} \right)^{-U /(\pi \Gamma)+{\mathcal O}(U^2)}&& \tau \ll |\epsilon| \ll \Gamma
\end{array}\right. ,
\label{tauren}
\end{align}
where we have defined $\tau^{\rm ren} = \tau + \Sigma^{{\rm M}, \Lambda=0}_{12}=\tau + \Sigma^{{\rm M}, \Lambda=0}_{23}$. The renormalization of the onsite energy $\epsilon$ of site 2 is of order $U^2$; the self-energy matrix elements $\Sigma^{\rm M}_{11}$ and $\Sigma^{\rm M}_{33}$ appear in the Green function in combination with $\Gamma$ and can thus be neglected in the scaling limit.

The renormalization of the hopping amplitude manifests physically in the charge susceptibility 
\begin{equation}
\chi=\left.\frac{d\bar n_{2}}{d\epsilon}\right|_{\epsilon=0} = - \frac{2}{\pi T_K}~,
\end{equation} 
which in turn is governed by the (renormalized) energy scale $T_K$. The latter is the universal scale of the model in equilibrium.\cite{Schlottmann} We will thus use it as our characteristic energy scale. The first-order FRG approximation to $T_K$ can easily be computed numerically from $\chi$.\cite{Karrasch10} 

\emph{The nonequilibrium steady-state physics} can again be solely attributed to a renormalization of the $1,2$ and $2,3$ matrix elements of the self-energy (with $\tau^{\rm ren}_{12} \neq \tau^{\rm ren}_{23}$ in presence of a finite bias voltage $V$). For $|\epsilon|, \tau \ll V$ the RG flow is cut off by $V$ and hence 
\begin{align}
\frac{\tau^{\rm ren}_{12}}{\tau}, \frac{\tau^{\rm ren}_{23}}{\tau} \sim  
\left( \frac{V}{\Gamma} \right)^{-U /(\pi \Gamma)+{\mathcal O}(U^2)} .
\label{taurenV}
\end{align} 
This leads to a power-law suppression of the current at large $V$:\cite{Karrasch10,Doyon,Schmitteckert,Andergassen} 
\begin{equation}
\frac{J_{L/R}}{\Gamma}  \sim \left(\frac{V}{\Gamma} \right)^{-2 U /(\pi \Gamma)+{\mathcal O}(U^2)} .
\end{equation}
Even richer physics can be observed (i) at $\epsilon=\pm V/2$ where the differential conductance exhibits a resonance peak, and (ii) for left-right asymmetric tunnel couplings to site 2.\cite{Karrasch10a,Andergassen} In both cases, the bias voltage no longer simply acts as an infrared cutoff. We do not consider these parameter regimes in the present paper and thus refrain from discussing further details.   

After this brief summary of the IRLM's scaling-limit physics, we now employ our newly-developed FRG scheme to study its real-time evolution. The flow equation \eqref{eq:FlowRet} takes the explicit form
\begin{align}
&\partial_\Lambda \Sigma^{\Ret}(t',t)=-i\frac{U}{2}\delta(t'-t)\notag\\
&\times\begin{pmatrix}
S_{22}^\K(t,t)&-S_{12}^\K(t,t)&0\\-S_{12}^\K(t,t)^*&S_{11}^\K(t,t)+S_{33}^\K(t,t)&S_{23}^\K(t,t)\\
0&S_{23}^\K(t,t)^*&S_{22}^\K(t,t)
\end{pmatrix} . 
\label{flowcomplete}
\end{align}
We again emphasize that computing the right-hand side of this set of differential equations requires the 
solution of both Eq.\ (\ref{eq:SK_trunc}) as well as the Dyson equation(s) (\ref{eins}) and (\ref{eq:GK_contrib}). We did not succeed in doing this analytically (in contrast to the steady-state limit where even the flow equation itself was amenable to an analytic treatment\cite{Karrasch10a}) and thus resort to a numerical solution. \textit{In principle}, the resulting set of coupled equations can be coded trivially. \textit{In practice}, however, the computational effort of a straightforward implementation is vast, and no results can be obtained in reasonable time. We thus devise an efficient algorithm to speed up numerics but do not apply further approximations (as does Ref.\ \onlinecite{PeterK}). The reader which is not interested in those computational details may skip the next section -- only keep in mind that the algorithm we are about to introduce is numerically exact.


\section{Numerical algorithm}
\label{sec:num}

To solve the set of equations (\ref{flowcomplete}), (\ref{eq:SK_trunc}), (\ref{eins}), 
and (\ref{eq:GK_contrib}) numerically we first discretize the continuous time variable $t$. 
The number of time steps $M$ determines the number of coupled differential equations 
that are eventually to be integrated. We are mainly interested in the universal scaling limit of the IRLM where the coupling to the 
reservoirs $\Gamma$ is much larger than the hopping amplitude $\tau$ of two adjacent dot sites. This naturally leads to two time regimes characterized by the 
scales $ 1/\Gamma \ll 1/\tau$. Both need to be resolved in order to access all times from transient to asymptotic. We thus need to employ small discretization steps ($\ll 1/\Gamma$) for small times. However, numerical resources limit $M$ to about $10^2$ (see below), so the step size necessarily increases for larger times such that eventually the scale $1/\tau$ on which the relaxation process towards the steady state takes place can be reached (this will be further specified below).

For discrete time variables $t_n$ (we set $ t_0=0 $) the full retarded Green function can be decomposed via Eq.\ \eqref{eq:multiformula}:
\begin{equation}
\begin{split}
\label{eq:Gretdecomposed}
G^{\Ret}(t,t')=&\left[\prod\limits_{n=1}^{n_t-(m+1)} G^{\Ret}(t_{n_t-n+1},t_{n_t-n})i\right]\\&\times G^{\Ret}(t_{m+1},t')\;\;\;\;\;\;\;\;\; \forall t>t' ,
\end{split}
\end{equation}
where the product extends over all (discrete) $ t_n $ from $ t_{n_t}=t $ to the first time where $ t_m<t' $. Thereafter, the time arguments of the retarded Green functions differ at most by one discretization step, and if the latter is chosen small enough, the self-energy can be approximated as piecewise constant (which is obviously numerically exact in the limit $M\to\infty$). This in turn allows evaluation of each retarded Green function in Eq.\  \eqref{eq:Gretdecomposed} via Eq.\ \eqref{eq:GRtdependent}:
\begin{equation}
\begin{split}
\label{eq:GRetdisc}
&G^{\Ret}(t_{n+1},t_{n}) = - i  e^{- i 
    \left[h_0^{\rm dot} +\tilde \Sigma^\Ret_{\res}-i\Lambda+\tilde \Sigma^\Ret_{\bar t}\right]\left[t_{n+1}-t_{n}\right] } ,
\end{split}
\end{equation}
where we introduced $\Sigma=\tilde\Sigma\delta(t)$, and $h_0^{\rm dot}$ denotes the single particle matrix 
representation of Eq.\ \eqref{eq:singlepartinter}. The index $\bar t$ of $\tilde \Sigma^\Ret$ indicates the mean value of the retarded self-energy in the small 
interval $[t_{n},t_{n+1}]$. The advanced Green function $G^\Adv$ is related to $G^\Ret$ through Eq.\ (\ref{adjoint}).

A similar argument applied to the Keldysh component yields the recursion relation
\begin{equation}
\label{eq:GKdis}
\begin{split}
&G^{\K}(t_{n+1} ,t_{n+1})\\
&=G^{\Ret}(t_{n+1},t_n)G^{\K}(t_n ,t_n)G^{\Adv}(t_n,t_{n+1})\\
&+\Bigg[-i\sum\limits_{m=0 }^{n-1}\;\;\int\limits_{t_{n}}^{t_{n+1}} ds_1\int\limits_{t_m}^{t_{m+1}} ds_2 G^{\Ret}(t_{n+1},s_1)\Sigma^\K_\res(s_1,s_2) \\
&\times G^{\Adv}(s_2,t_{m+1}) G^{\Adv}(t_{m+1},t_{n+1})-\text{H.c.}\Bigg]\\
&+\int\limits_{t_n}^{t_{n+1}} ds_1\int\limits_{t_n}^{t_{n+1}} ds_2 G^{\text{ret}}(t_{n+1},s_1)\Sigma^\K_\res(s_1,s_2)  
G^{\text{adv}}(s_2,t_{n+1}) ,
\end{split}
\end{equation} 
with the initial condition $G^\K(0,0)=-i(1-2 \bar n)$. We have exploited that $G^\K$ and $ S^\K$ only enter the right-hand side of the flow equation (\ref{flowcomplete}) with equal time arguments. The same holds if we want to compute observables: For the occupancy $\bar n_i(t)$ this is apparent from Eq.\ (\ref{occu}); for the current $J_\alpha(t)$ of Eq.\ (\ref{curr}) it follows from time locality of $\Sigma_\alpha^\Ret$ in the wide band limit. The integrals in Eq.\ \eqref{eq:GKdis} can be further evaluated and ultimately expressed in terms of exponential integrals. This is discussed in the Appendix [see Eqs.\ (\ref{eq:GK3sites1}) and (\ref{eq:GK3sites1_a})]. Finally, $ S^{\K}(t_{n+1} ,t_{n+1}) $ can be obtained from an analogous procedure.

The remaining differential (flow) equation (\ref{flowcomplete}) can be implemented straightforwardly using standard Runge-Kutta routines. Computing the right-hand side through the above recursive procedure scales as $M^2$, and thus calculating $\bar n_i(t)$ and $J_\alpha(t)$ for a given parameter set approximately 
scales as $M^3$. We carefully ensure to choose the numerical control parameter $M$ large enough for our results to be numerically exact (typically $M\sim10^2$ is sufficient;\cite{commentnumpara} numerics can be carried out in reasonable time on standard PCs). Thus, no additional approximation -- such as the often applied\cite{PeterK} but in many applications uncontrolled generalized Kadanoff-Baym ansatz \cite{HaugJauho} -- is used to integrate the Dyson equation.

Solving equation \eqref{eq:GKdis} is particularly simple at $U=0$. Since the self-energy associated with the two-particle interaction vanishes, there is no need to discretize time. One solely has to determine the last term of Eq.\ \eqref{eq:GKdis} with $ t_{n+1} $ and $ t_n $ replaced 
by $ t $ and $ 0 $, respectively (the recursion becomes trivial). The remaining integral can be expressed in terms of exponential integrals [the result is given by Eq.\ \eqref{eq:GK3sites1} with $ t_{n+1}\to t $, $ t_n\to0 $ as well as $ \Lambda $, $\Sigma^\Ret_{\bar t}$, and $U$ set to zero]. From $G^{0,\K}(t,t)$ one can calculate the noninteracting occupancy $\bar n_i(t)$ through Eq.\ (\ref{occu}) and the current $J_\alpha(t)$ via Eq.\ (\ref{curr}). We will discuss our results for $U=0$ in Sec.\ \ref{sec:resultsnoint}.


\section{Time evolution in the IRLM}
\label{sec:results}

\subsection{The noninteracting case} 

\label{sec:resultsnoint}

\begin{figure}[t]
   \centering \includegraphics[width=0.9\linewidth,clip]{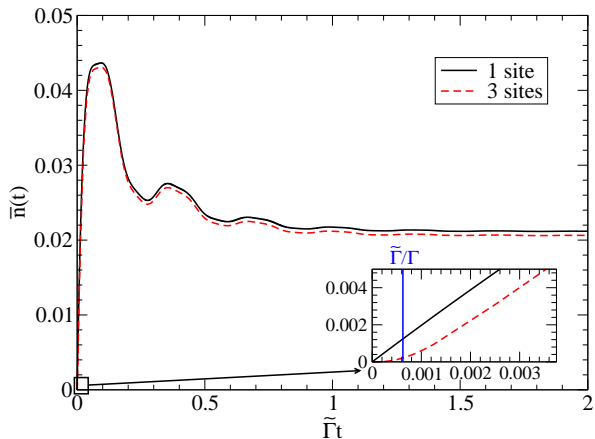}
\caption{(Color online) Comparison of the central-site occupancy between a three-site and a (field-theoretical-like) one-site version of the interacting resonant level model for scaling-limit parameters $\epsilon/\Gamma=V/\Gamma=0.025$, and ${\tau}/{\Gamma}=0.025$ (the hybridization in the single-site model reads $\tilde \Gamma=\tau^2/\Gamma$; see the main text for details). The inset illustrates that both models differ only for times $ \sim 1/ \Gamma $.}
   \label{fig:1vs3dot}
\end{figure}

As an instructive step towards an understanding of the nonequilibrium relaxation dynamics of the IRLM we study the case $U=0$. Throughout this section (and for the whole rest of the paper) we assume that the three dot sites are empty at $t=0$ but stress that other initial states can be considered without any additional effort (as long as they fulfill the criteria mentioned in Sec.\ \ref{sec:Keldysh_formalism}). We will first quantify the heuristic (yet reasonable) statement that in the scaling limit defined by Eq.\ \eqref{eq:Scaling limit} our three-site model becomes equivalent to the field-theoretical-like version of the IRLM where a single site couples to two reservoirs. Parameters can obviously be fixed by choosing
\begin{equation}
\frac{\tau^2}{\Gamma} = \tilde\Gamma
\end{equation} 
for a given hybridization strength $\tilde\Gamma$ of the single-site model. The latter can be solved analytically at $U=0$; exact expressions for the time-evolution of $\bar n$ and $J_\alpha$ for an initially empty and decoupled dot can be found in Ref.\ \onlinecite{Andergassen} (see also Ref.\ \onlinecite{Anders2}).\cite{fehler} In Fig. \ref{fig:1vs3dot} the occupancy number is compared to $\bar n_2(t)$ in our three-site dot (the parameters are given by $\tau/\Gamma=\epsilon/\Gamma=V/\Gamma=0.025$). The two models show different behavior at times smaller than $1/\Gamma$ (see the inset of Fig.\ \ref{fig:1vs3dot}), which is the time needed to fill the initially empty sites 1 and 3 of the three-site model to their steady state values. In the latter the displacement current fills the 1 and 3 site first, whereas in the single-site model it gives rise to a nonvanishing first derivative of $\bar n$ for $t\to 0$.\cite{Karrasch10a,Andergassen} The difference of the occupancy at small times leads to an offset at larger times (which however vanishes in the extreme scaling limit $\Gamma \to \infty$). Otherwise both models yield identical results.

For times $t \gg 1/\tilde \Gamma$ the occupancy and the current in the single-site model are given by (note that $T_K=4\tilde\Gamma$ at $U=0$)\cite{Andergassen}
\begin{align}
\bar n (t)\approx &\bar n_{\text{stat}}(1+e^{-4\tilde \Gamma t})+\frac{2}{\pi} \tilde \Gamma e^{-2 \tilde \Gamma t}
\notag\\&\times\left(\frac{\sin[(\epsilon-V/2)t]}{(\epsilon-V/2)^2t}+\frac{\sin[(\epsilon+V/2)t]}{(\epsilon +V/2)^2t}\right) , \label{nlarget}\\
\frac{J_\alpha (t)}{\tilde \Gamma }\approx&1-2 \bar n (t)+\frac{2}{\pi} \bigg(-\arctan\left[
\frac{\epsilon-\mu_\alpha}{2 \tilde \Gamma }\right]\notag\\
&-e^{-2 \tilde \Gamma  t}{\rm Im}\left[ \frac{e^{-i(\epsilon-\mu_\alpha) t}}{(i(\epsilon-\mu_\alpha)+2 \tilde \Gamma)t}
\right]\bigg) ,
\label{currentlarget}
\end{align}
where we have introduced the stationary occupancy 
\begin{align}
\bar n_{\text{stat}}=\frac{1}{2}+\frac{1}{2\pi}\sum\limits_{\alpha}\arctan\left(
\frac{\mu_\alpha-\epsilon}{2 \tilde \Gamma}\right) . 
\end{align}
The corresponding expressions of the three-site model at large $\Gamma$ are identical (if $\bar n$ and $\tilde\Gamma$ are replaced by $\bar n_2$ and $\tau^2/\Gamma$, respectively). Equation\ (\ref{nlarget}) illustrates that the long-time behavior of the occupancy exhibits two {\it oscillatory} terms with frequencies $\epsilon\pm V/2$ as well as an {\it exponential relaxation} with rates $ 4 \tilde \Gamma $ {\it and} $ 2 \tilde \Gamma $; the 
latter is accompanied by a {\it power-law correction} $1/t$. The current characteristics are similar. For later reference we note that in the limit $ |\epsilon\pm V/2|\gg T_K $ the current $ J_\alpha $ only shows the single frequency $ \epsilon-\mu_\alpha $ since the second frequency (originating from the occupancy term only) is suppressed.
 
\subsection{The interacting case}

\label{sec:resultsint}

After these prerequisites we now turn to the interacting case. Our objective is twofold. We first discuss our results from a pure physical perspective, \textit{assuming} that our FRG scheme yields accurate data as long as $U$ is not too large. This is certainly reasonable in light of the method's success to capture correlation effects within the IRLM in equilibrium and steady-state nonequilibrium.\cite{Karrasch10}  Thereafter, we relate our findings to real-time RG calculations (Sec.\ \ref{sec:resultsrtrg}). The RTRG access to nonequilibrium dynamics differs strongly from the FRG approach, and so do the applied (controlled) approximations. Thus, comparing with RTRG data provides a highly nontrivial test for the newly-developed FRG framework. 

\begin{figure}[t]
\centering
\includegraphics[width=0.9\linewidth,clip]{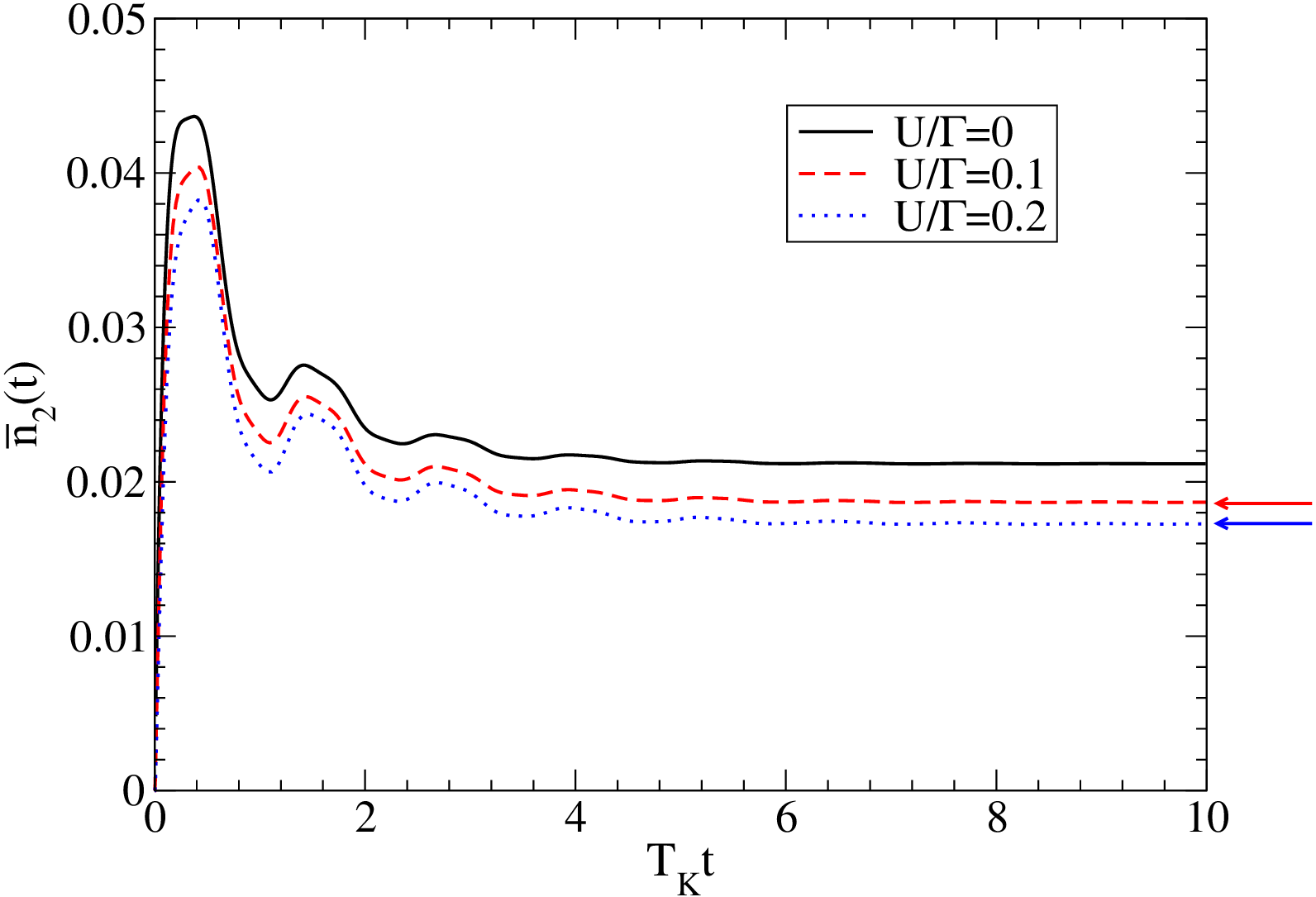}  
 \caption{(Color online) FRG data for the time evolution of the central-site occupancy of the interacting resonant level model depicted in Fig.\ \ref{fig:irlm}. The parameters read $\tau/\Gamma=0.025$ (local coupling), $\epsilon/T_K=10$ (level position), and $V/T_K=10$ (difference of chemical potentials). The universal equilibrium energy scale $T_K$ is renormalized through the local Coulomb interaction $U$. The arrows indicate the steady-state values obtained by the nonequilibrium steady-state functional RG of Ref.\ \onlinecite{Karrasch10}.} 
   \label{fig:Timeevolutionocc}
\vspace{.5cm}
\centering
 \includegraphics[width=0.9\linewidth,clip]{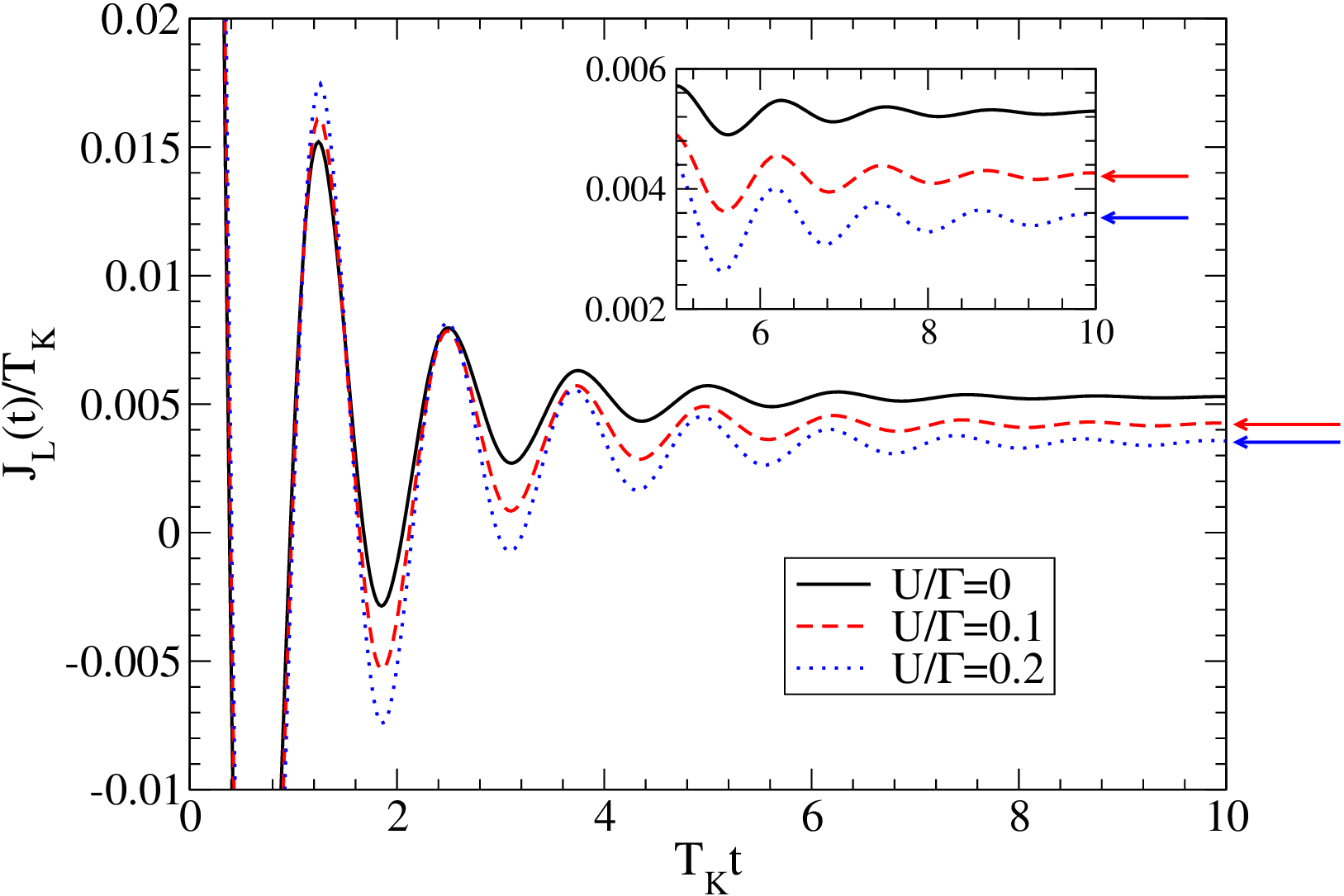}
\caption{(Color online) The same as in Fig.\ \ref{fig:Timeevolutionocc} but now showing the current leaving the left reservoir. The negative current found at small times was already observed in Refs.\ \onlinecite{Karrasch10a} and \onlinecite{Andergassen}.} 
   \label{fig:Timeevolutioncur}
\end{figure} 

Our discussion of the interacting case focusses exclusively on the limit $ |\epsilon\pm V/2|\gg T_K $, which is \textit{a posteriori} motivated by the fact that the physics can be interpreted in a simple way. Moreover, the far-from-equilibrium case $V\gg T_K$ is undoubtedly most intriguing on general grounds. We again point out that the FRG is not bound to any parameter regime -- it can flexibly describe the whole crossover from $V\ll T_K$ to $V\gg T_K$ (to $V\gg\Gamma$).

The results for the time dependence of the occupancy $ \bar n_2(t) $ and the current $ J_L(t) $ at different $U$ but fixed $\tau/\Gamma=0.025$, $\epsilon/T_K=10=V/T_K$ are depicted in Figs.\ \ref{fig:Timeevolutionocc} and \ref{fig:Timeevolutioncur}. The axes are scaled using the universal equilibrium energy $T_K$. Note that $\epsilon$ and $V$ need to be varied with $U$ in order to keep $\epsilon/T_K$ and $V/T_K$ fixed. Whereas the steady-state values of $\bar n_2$ and $J_L$ are nonuniversal, the oscillation frequencies depend only very weakly on the interaction strength $U$ (the relative position of maxima and minima remain unaltered). In analogy to the noninteracting case, only a single frequency $|\epsilon+V/2|$ governs the current in the limit $ |\epsilon\pm V/2|\gg T_K $ while both $|\epsilon\pm V/2|$ are manifest in the occupancy.

\begin{figure}[t]
\centering  
   \includegraphics[width=0.9\linewidth,clip]{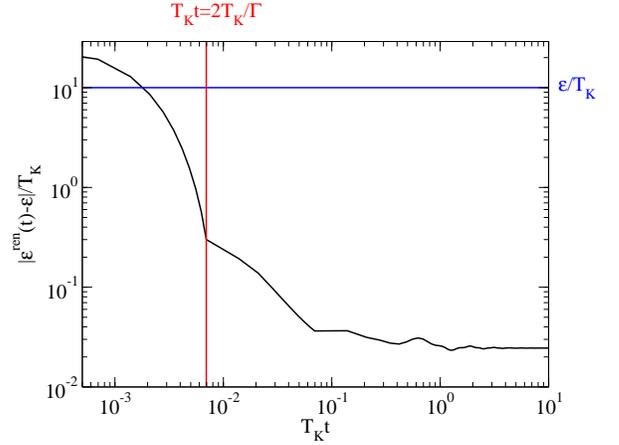}
   \caption{(Color online) Time dependence of the renormalized onsite energy $ \epsilon^{\rm ren}(t)-\epsilon \in {\mathbb R}$ (of the central site 2). The parameters are the same as in Fig.\ 
\ref{fig:Timeevolutionocc}: $\tau/\Gamma=0.025$, $\epsilon/T_K=V/T_K=10$, and $ U/\Gamma=0.1$. Note the double logarithmic scale. Kinks in the graph are due to the time discretization.} 
   \label{fig:renormeps}
\end{figure}

\begin{figure}
\centering  
   \includegraphics[width=0.83\linewidth,clip]{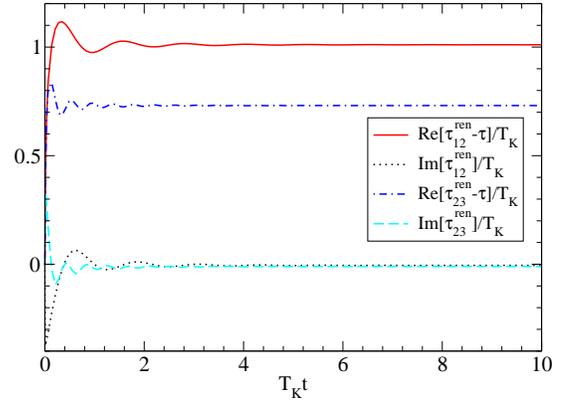}
\caption{(Color online) The same as in Fig.\ \ref{fig:renormeps} but for the renormalized hoppings $ \tau_{12}^{\rm ren} - \tau = \tilde \Sigma^{\Ret,\Lambda=0}_{12} \in {\mathbb C}$ between sites 1 and 2, 
and $ \tau_{23}^{\rm ren} - \tau = \tilde \Sigma^{\Ret,\Lambda=0}_{23} \in {\mathbb C}$ between sites 2 and 3.}
   \label{fig:renormth}
\end{figure}

It is a distinct advantage of the first-order FRG framework that it allows to gain intuitive physical insights from investigating the (time dependence of the) renormalized effective single particle parameters. Figure\ \ref{fig:renormeps} shows the renormalized onsite energy $ \epsilon^{\rm ren}(t)-\epsilon = \tilde \Sigma^{\Ret,\Lambda=0}_{22} \in {\mathbb R}$ (note the double logarithmic scale). For times larger than $1/\Gamma$, it does not flow to leading order in $U$. The renormalized hopping amplitudes $\tau_{12}^{\rm ren} - \tau = \tilde \Sigma^{\Ret,\Lambda=0}_{12} \in {\mathbb C}$ and $ \tau_{23}^{\rm ren} - \tau = \tilde \Sigma^{\Ret,\Lambda=0}_{23} \in {\mathbb C}$ are depicted in Fig.\ \ref{fig:renormth} (recall that $\tau_{12}^{\rm ren}=\tau_{23}^{\rm ren}$ only in absence of a bias voltage). They begin to oscillate very quickly around their steady-state values. The renormalized hopping between sites 1 and 2 (between 2 and 3) only exhibits the smaller frequency $ \epsilon-V/2 $ (the larger $ \epsilon+V/2 $), which was checked via Fourier transformation. Thus, the oscillation frequencies $ \epsilon\pm V/2 $ which govern observables at $U=0$ are not altered through the Coulomb interaction; the latter merely leads to $U$-dependent phase shifts.


\begin{figure}
\centering  
   \includegraphics[width=\linewidth,clip]{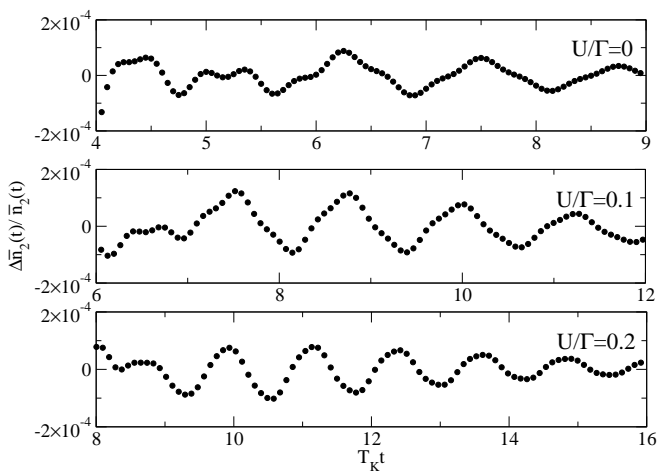}
\caption{Relative difference of the data for the occupancy shown in 
Fig.\ \ref{fig:Timeevolutionocc} and a fit to  Eq.\ (\ref{fitform}). Small deviations even in the noninteracting case originate from $ \Gamma $ being finite and t being only roughly an order of magnitude larger than $ T_K $.}
   \label{fig:ResdiffU}
\end{figure}

This observation indicates that the time evolution of $\bar n_2$ and $J_\alpha$ in the interacting case might be described by an analytic expression similar to the one at $U=0$ if the latter is supplemented by $U$-dependent prefactors, rates, and phase shifts. For a further investigation we make the ansatz (sticking to $t \gg 1/\tilde \Gamma$ for simplicity)
\begin{equation}
\begin{split}
\bar n_2(t)=a_0+a_1e^{-2a_2t}&+a_3e^{-a_4t}
\frac{\sin\left[ \left( \epsilon^{\rm ren}-\frac{V}{2}+\phi \right) t \right]}
{\left(\epsilon^{\rm ren}-\frac{V}{2}\right)^2t}\\
&+a_5e^{-a_4t}\frac{\sin\left[\left(\epsilon^{\rm ren}+\frac{V}{2}+\phi \right)t \right]}
{\left(\epsilon^{\rm ren}+\frac{V}{2}\right)^2t} .
\label{fitform}  
\end{split}
\end{equation}
This is precisely the form of Eq.\ (\ref{nlarget}) if it is generalized to account for left-right asymmetric bare hoppings ($\tau_{12} \neq \tau_{23}$ is required at $ V> 0 $); one can easily show that at $U=0$, the coefficients are then given by $a_0=a_1= \bar n_{2,\text{stat}}(U=0)$, $a_2=a_4=\tilde \Gamma_{12} + \tilde \Gamma_{23}$, $a_3=2 \tilde \Gamma_{12} /\pi $, and $a_5=2 \tilde \Gamma_{23} /\pi $, where $\tilde \Gamma_{ij} = \tau_{ij}^2/\Gamma$. At $U>0$, the $a_i$ serve as fit parameters. Our conjecture that a crucial part of the interaction effects can indeed be incorporated by taking the noninteracting functional form but renormalized parameters is then strongly supported by the fact that (i) the fitting error \textit{does not increase with $U$} (see Fig.\ \ref{fig:ResdiffU} for an illustration), and (ii) the interpretation of the coefficients is consistent with the noninteracting case: $a_0$ and $a_1$ are very close to the $U$-dependent steady-state occupancy as directly obtained by steady-state functional RG\cite{Karrasch10} or averaging the real-time data for large times; likewise, the rates $a_{2,4}$ and the asymmetry ratio $a_3/a_5$ agree with the asymptotes of the renormalized $\tilde \Gamma_{12}$, $\tilde \Gamma_{2,3}$, and $\tilde \Gamma_{12}/\tilde \Gamma_{2,3}$ (see Tab.\ \ref{Fitparasmore}). In passing, we note that the phase $\phi$ increases from  $\phi(U/\Gamma=0)=0$ to $\phi(U/\Gamma=0.2) \approx 0.26$ for the parameters of Fig.\ \ref{fig:Timeevolutionocc} and thus shows a significant $U$-dependence.

\begin{table}[b]
\centering
\begin{tabular}{cccccc}
\vspace{0.1cm}
$ U/\Gamma $&$\;(\tilde \Gamma_{12}+\tilde\Gamma_{23})/T_K\;$&$\;\tilde\Gamma_{12}/\tilde\Gamma_{23}\;$&$a_2/T_K$&$a_4/T_K$&$ a_3/a_5 $\\
\vspace{0.1cm}
$0 $&$ 0.500 $&$ 1.000 $&$ 0.504\substack{{0.505}\\{0.502}} $&$ 0.496_{0.493}^{0.498} $&$ 1.056  $\\
\vspace{0.1cm}
$0.1 $&$ 0.453 $&$ 1.073 $&$ 0.463_{0.458}^{0.467} $&$ 0.445_{0.440}^{0.451} $&$ 1.071  $\\
\vspace{0.1cm}
$0.2 $&$ 0.417 $&$ 1.149 $&$ 0.389^{0.398}_{0.381} $&$ 0.405_{0.398}^{0.413} $&$ 1.167 $\\
\end{tabular}\\
\caption{The fitting parameters $ a_{2/4} $ in comparison to the renormalized relaxation rates $ \tilde\Gamma_{ij}=\frac{|\tau^\ix{ren}_{ij}|^2}{\Gamma} $ for figure \ref{fig:Timeevolutioncur}. The error is with respect to a $ 68\% $ confidence level.} 
\label{Fitparasmore}
\end{table}

\subsection{Comparison to real-time RG}

\label{sec:resultsrtrg}

In this section we relate the data from our newly-developed real-time FRG scheme to independently obtained results. As a first consistency check we compare the long-time asymptotes of the current and the occupancy with a direct calculation using lowest-order nonequilibrium steady-state functional RG.\cite{Karrasch10} As illustrated in Figs.\ \ref{fig:Timeevolutionocc} and \ref{fig:Timeevolutioncur} -- where arrows indicate values extracted from the independent steady-state framework -- the agreement is quantitative.

The time evolution of the IRLM exposed to a bias voltage was recently studied using real-time RG.\cite{Karrasch10a,Andergassen} The RTRG is based on approximations that are controlled at small $U$ \textit{and} $\tau$. In equilibrium and steady-state nonequilibrium FRG and RTRG data was shown to agree on a quantitative level at small $U$.\cite{Karrasch10a} For the discussion below one should bear in mind that it is not straightforward within the RTRG to obtain prefactors correctly to order $U$ due to technical (truncation) subtleties.\cite{Karrasch10a,Andergassen} Rates and exponents of possible power-law corrections, however, come out correctly to order $U$.\cite{Andergassen} This is different in the truncated FRG used here where it is guaranteed that the prefactors are correct to leading order in the interaction. 

\begin{figure}[t]
\centering
\includegraphics[width=\linewidth,clip]{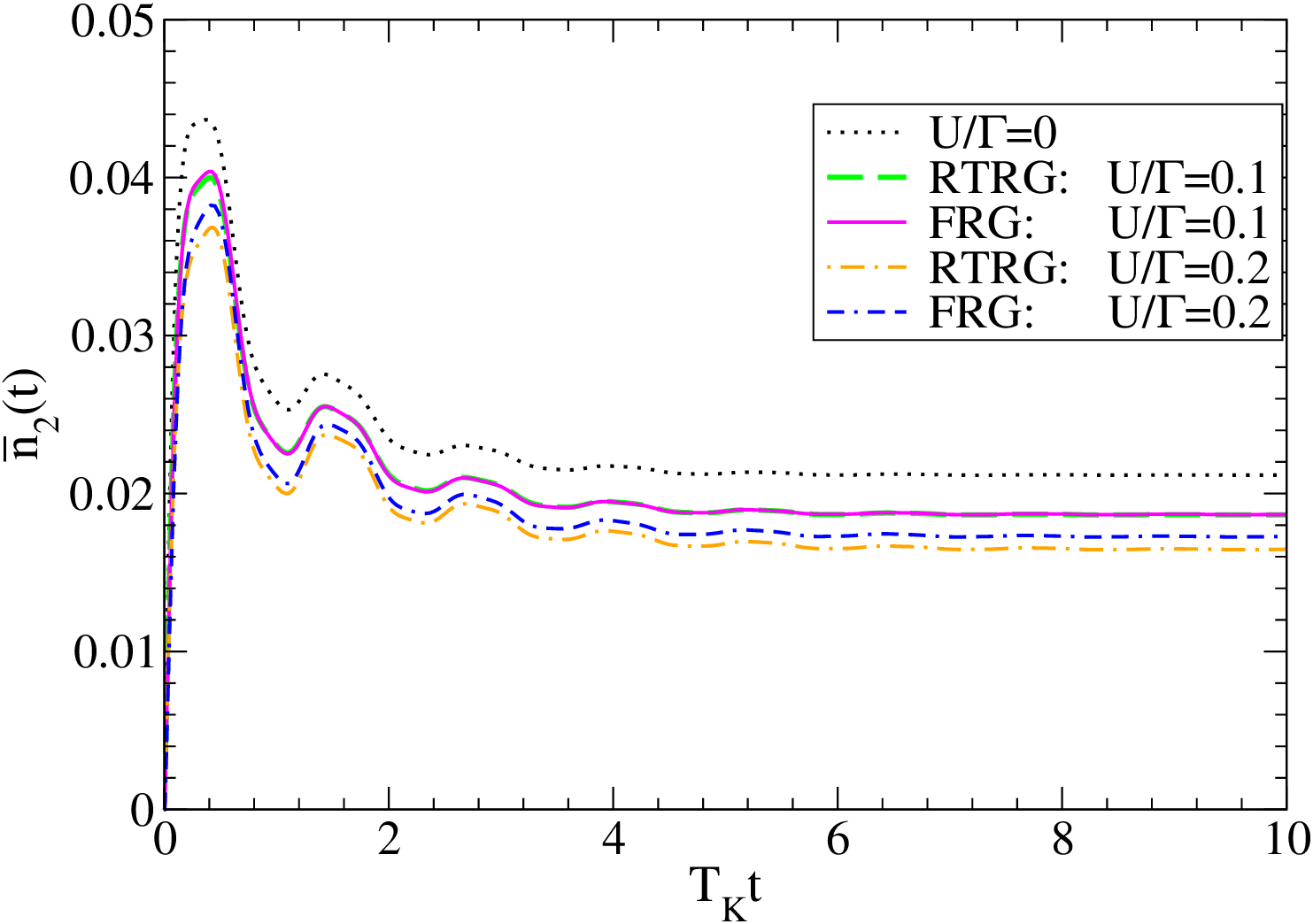}  
\caption{(Color online) Comparison of FRG and real-time RG data for the occupancy $ \bar n_2(t) $ at $\tau/\Gamma=0.025$, and $\epsilon/T_K=V/T_K=10$ (the parameters of Fig.\ \ref{fig:Timeevolutionocc}). The curves at $ U/\Gamma=0.1 $ are almost indistinguishable.} 
   \label{fig:RTRGvsFRGocc}
\vspace{.5cm}
\centering
\includegraphics[width=\linewidth,clip]{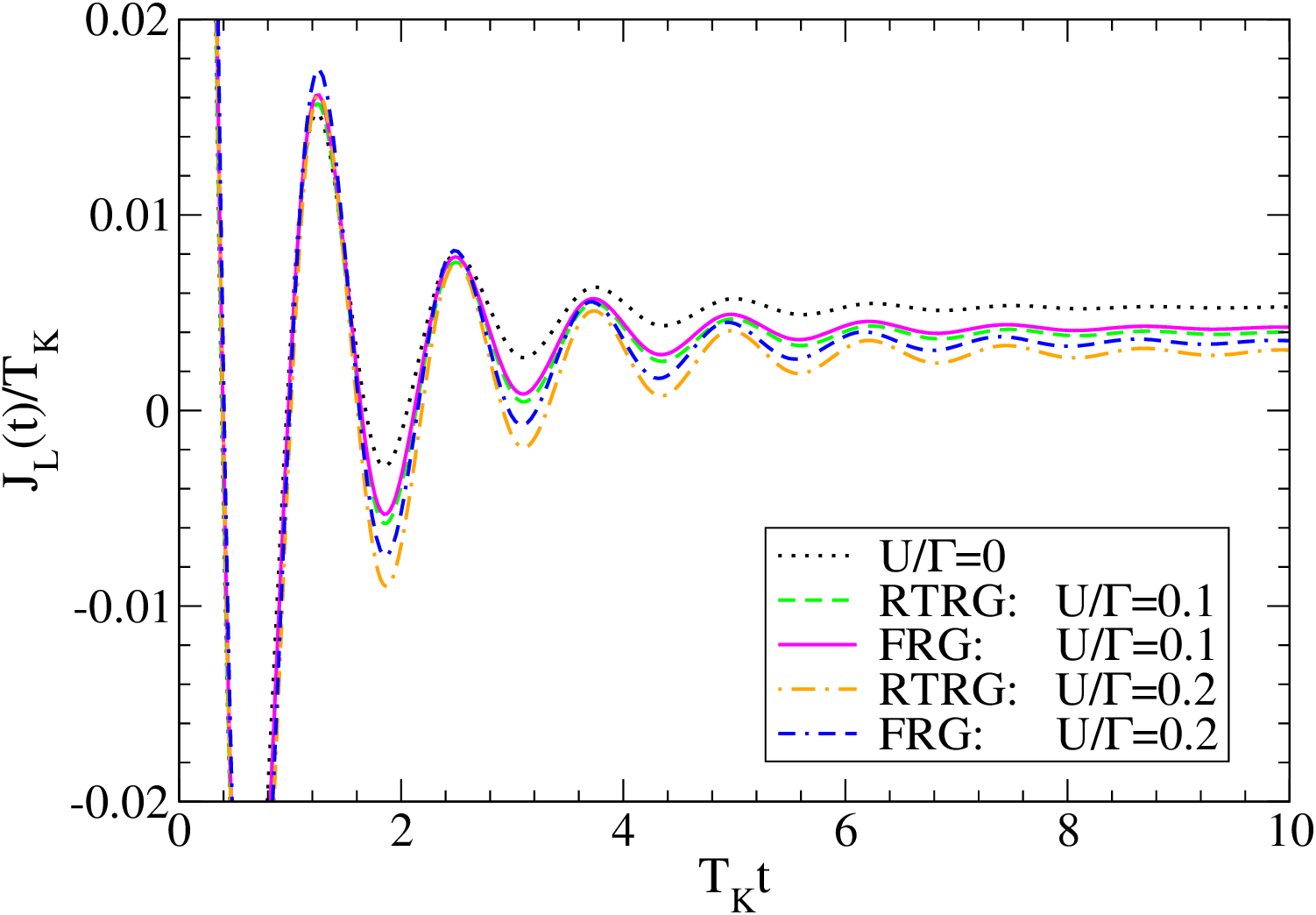}
\caption{(Color online) The same as in Fig.\ \ref{fig:RTRGvsFRGocc} but for the current.} 
   \label{fig:RTRGvsFRGcur}
\end{figure}

A comparison of the occupancy and the current obtained by FRG/RTRG is shown in Figs.\ \ref{fig:RTRGvsFRGocc} and \ref{fig:RTRGvsFRGcur}; the agreement is satisfying. Small differences may be attributed to the fact that our parameter set is not in the extreme scaling limit (for a quantitative analysis of this, see Sec.\ \ref{sec:resultsnoint}), while the real-time RG is directly set up in this regime. Deviations of similar magnitude were observed in prior steady-state calculations.\cite{Karrasch10} Moreover, RTRG prefactors are not controlled to leading order, and this issue becomes more relevant at larger interactions.

It is a key advantage of the real-time RG that one can derive approximate analytical expressions for the long time behavior of the occupancy and the current 
in the regime $ |\epsilon\pm V/2|\gg T_K $ even in presence of (not too large) interactions.\cite{Karrasch10a,Andergassen} In the unbiased case, the current at sufficiently large $t$ reads (observe that $V=0$ implies $J(t\to\infty)\to0$)
\begin{equation}
J_L(t)=a_1e^{-\bar \Gamma t}+\frac{T_K}{2\pi}(T_Kt)^{g}
e^{-\bar \Gamma_\epsilon t/2}\frac{\cos(\bar \epsilon t)}{\bar \epsilon t},\label{eq:Ifit}
\end{equation}
where $ \bar\Gamma $, $ \bar\Gamma_\epsilon $, and $\bar \epsilon$ denote the RTRG-renormalized decay rates and level position. One particularly finds an interaction-dependent power-law correction $1/t^{1-g}$, with $g=2U/\pi + {\mathcal O} (U^2)$ [compare to Eq.\ (\ref{currentlarget}) for $U=0$]. Real-time RG predicts power-law corrections also for $V>0$.\cite{Karrasch10a,Andergassen} 
 
It is an important test for our approximate functional RG approach -- which is based on a truncation scheme guided by perturbation theory in orders of $U$ -- to investigate whether the RG resummation not only leads to an exponential decay in time with $U$-dependent decay rates (as was already proven above) but whether it {\it simultaneously} gives rise to power laws in $t$ with $U$-dependent exponents. The predicted power-law corrections to some of the exponential decay terms\cite{Karrasch10a,Andergassen} are subleading corrections. Achieving a rather good fit of our functional RG data with the ansatz of Eq.\ (\ref{fitform}) thus does not exclude the presence of such terms.   
 
\begin{figure}[t]
\centering
\includegraphics[width=0.9\linewidth,clip]{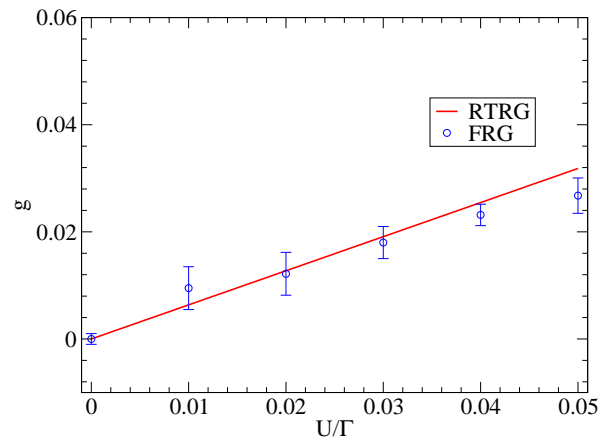}
\caption{Interaction dependence of the power-law exponent extracted from our FRG data by fitting to the RTRG prediction of Eq.\ (\ref{eq:Ifit}) at $\epsilon/T_K=30$ (see the main text for details). The solid line shows the leading order term $2 U/\pi$. The error bars refer to a $68\%$ confidence level.} 
   \label{fig:power-law}
\end{figure}

We now analyze our FRG data for potential power-law contributions. For reasons of simplicity, we focus on the unbiased case where the hopping amplitudes renormalize symmetrically. Note that due to the structure of Eq.\ \eqref{eq:Ifit} a fitting procedure can easily shift the influence of a power law to a small variation of the relaxation rates. We thus do {\em not} fit the relaxation rates but rather take them as their steady-state renormalized values (see Sec.\ \ref{sec:resultsint}, in particular Fig.\ \ref{fig:renormth} and Tab.\ \ref{Fitparasmore}). Likewise, we replace $\bar \epsilon$ by our steady-state renormalized level position (see Fig.\ \ref{fig:renormeps}). Thereafter, we fit the FRG data to the form predicted by real-time RG [Eq.\ \eqref{eq:Ifit}] employing $a_1 $ and $g$ as fitting parameters. We emphasize that the prefactor $T_K$ of the oscillatory term in  Eq.\ \eqref{eq:Ifit} does not suffer from the above mentioned subtleties of the RTRG, but $a_1$ does and must thus be fitted. As illustrated in Fig.\ \ref{fig:power-law}, the $U$-dependence of the extracted exponent $g$ agrees with the real-time RG prediction within the error bars. This indicates that our functional RG indeed achieves a resummation which simultaneously leads to an exponential time decay as well as {\it power-law corrections}, both featuring $U${\it -dependent exponents.} Verifying this analytically (e.g., using the cutoff introduced in Ref.\ \onlinecite{Gasenzer}) will be subject of a future investigation.

\section{Results for the ohmic spin-boson model}
\label{sec:spinboson}

As a second application we focus on a special parameter regime for which the IRLM can be mapped onto the so-called ohmic spin-boson model.\cite{spinbosonrmp} The latter constitutes one of the basic models used to study decoherence and relaxation phenomena in quantum systems, and its relaxation dynamics was previously investigated by a variety of methods, in particular field theory\cite{Saleur} and an improved noninteracting blip approximation.\cite{Egger} Relating our data to those results provides an additional test for the newly-developed time-dependent functional RG approach.

The spin-boson model describes two fermionic states tunnel-coupled by $\Delta$ and separated by an energy $E$ which interact with a bath of bosons:
\begin{equation}
H_{\text{SB}}=\frac{E}{2}\sigma_z -\frac{\Delta}{2}\sigma_x
+\sum\limits_q \omega_{q}a^\dagger_qa_q +\frac{\sigma_z}{2}\sum\limits_q g_q (a_q+a_q^\dagger),
\end{equation}
where $a_q^{(\dag)}$ are bosonic lowering (raising) operators, $\sigma_{x/z}$ denote the Pauli matrices, $ g_q $ describes the momentum $q$ dependent coupling to the phonon bath, and $\omega_q$ the dispersion of the latter.  For the so-called ohmic case where the spectral density reads
\begin{equation}
J(\omega)=\pi\sum\limits_q g_q^2\delta(\omega-\omega_q)=2\pi \alpha \omega e^{-\omega/\omega_c},
\end{equation}
with $\omega_c$ being a high-frequency cutoff, one can map the spin-boson model to the unbiased ($V=0$) IRLM.\cite{spinbosonrmp} The parameters are related through
\begin{align}
\epsilon &= E \\
U/(\pi \Gamma )&=(1-2\alpha)/2\\
\tilde \Gamma&= \frac{\tau^2}{\Gamma}=\frac{\Delta^2}{8\omega_c}.
\end{align}
The occupancy of the IRLM determines the expectation value of $\sigma_z$ via
\begin{equation}
\bar n(t) = \frac{\left< \sigma_z \right> +1}{2} , 
\end{equation}
where we can identify $\bar n(t) = \bar n_2(t)$ of our three site dot model in the scaling limit. We again assume an initially empty dot, i.e.\ a spin pointing in $-z$-direction at $t=0$. 

For $\epsilon=E=0$ the ohmic spin-boson model supposedly exhibits a coherent-incoherent transition at $\alpha=1/2$ (where the interaction $U$ of the IRLM changes from repulsive to attractive).\cite{spinbosonrmp} Field 
theory\cite{Saleur} and an improved noninteracting blib approximation\cite{Egger} both predict the relaxation rate and the frequency $\Omega$ (in the coherent phase) to be related through
\begin{equation}
\frac{\Omega}{4 \tilde \Gamma}= 2U/\Gamma + {\mathcal O}(U^2)~.
\label{eq:freqpredSP}
\end{equation}
We have dropped second-order terms which are not consistently included within our truncation scheme.

\begin{figure}[t]
\centering  
\includegraphics[width=0.9\linewidth,clip]{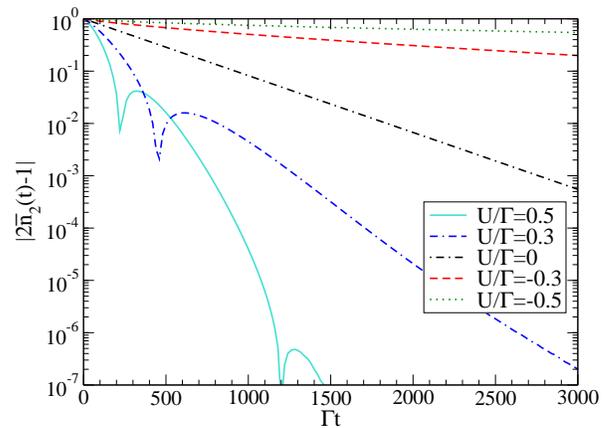}   
\caption{(Color online) FRG calculation of the expectation value $ |\left<  \sigma_z(t) \right> |=|2n_2(t)-1|$ for the ohmic spin-boson model. In absence of a bias voltage, the latter can be mapped to the IRLM, and our fermionic FRG scheme is directly applicable. In terms of the IRLM, the system parameters read $ \tau/\Gamma=0.025 $ and $ \epsilon=V=0$. Note that the $y$-axis is scaled logarithmically and that the times are given with respect to $ \Gamma $ instead of $ T_K $.} 
   \label{fig:SpinBoson_t}
\end{figure}

\begin{figure}[t]
\centering  
\includegraphics[width=0.9\linewidth,clip]{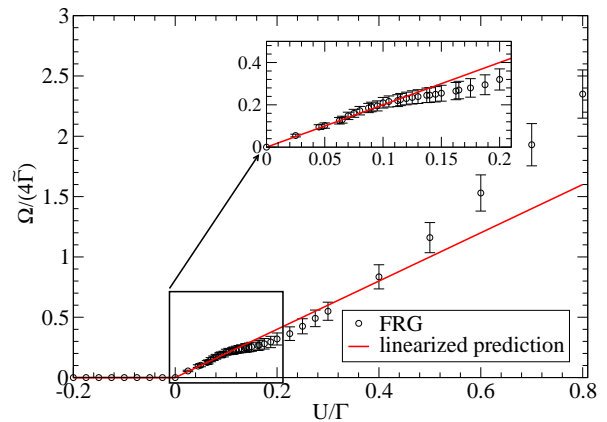}   
\caption{(Color online) Oscillation frequency governing the relaxation dynamics of the spin-boson model as a function of $U$ for $\tau/\Gamma=0.025$. The 
functional RG result is compared to the prediction of Eq.\ (\ref{eq:freqpredSP}). The error bars result from a $68\%$ confidence level fitting of the frequency.} 
   \label{fig:SpinBoson}
\end{figure}

FRG results for the time evolution of $ |\left< \sigma_z(t) \right>|=|2 \bar n_2(t)-1|$ are shown in Fig.\ \ref{fig:SpinBoson_t}. As expected, the exponentially damped oscillations (indicated by the dips) present for $U>0$ disappear at $U<0$ as the dynamics turn incoherent. We scale the $y$-axis logarithmically (i) to emphasize the exponential damping, and (ii) to illustrate the coherent-incoherent transition at $U=0$ more clearly. Note that only a single oscillation period manifests even for the largest interaction. This is qualitatively consistent with Eq.\ (\ref{eq:freqpredSP}) which postulates a frequency $4 U  \tilde \Gamma/\Gamma $; in the scaling limit, this is a very small number if $U$ is not too large. For a quantitative analysis, we fit our FRG data to the predicted long-time behavior:
\begin{equation}
\bar n_2(t)=0.5 + a e^{-\Gamma t} \cos(\Omega t)~.
\end{equation} 
The frequency $\Omega$ extracted from the FRG framework agrees with Eq.\ \eqref{eq:freqpredSP} even for fairly large values of $U$ (see Fig.\ \ref{fig:SpinBoson}). One should note that the coherent-decoherent transition is exclusively driven by the interaction (i.e., by the coupling of the spin to the bosons). Reproducing the transition as well as the relation between the relaxation rate and oscillation frequency (the latter being proportional to $U$!) in the coherent regime constitutes another stringent test of our approximate approach.

\section{Conclusion}
\label{sec:conclusion}

We developed a functional RG approach to study time-dependent 
electron transport through quantum dots with a few correlated degrees of 
freedom coupled to Fermi liquid reservoirs. It allows 
to investigate the relaxation dynamics out of an initial 
nonequilibrium state into the steady state driven by a bias voltage. Additionally, 
the dynamics of Hamiltonians with time-dependent parameters 
can be tackled. The hybridization flow parameter was used, and the hierarchy of functional RG flow-equations was truncated to lowest order.

We applied this approach to the interacting resonant level model with time independent 
parameters and computed the time dependence of the occupancy of the dot level 
and the current through it. We devised an efficient algorithm to solve the Dyson equation 
as a key step for a numerically exact implementation of the flow equations. The relaxation dynamics of the system exposed to a bias voltage 
is dominated by an exponential decay with two different rates and oscillations with two different frequencies. While 
the decay rates are significantly renormalized by the interaction, the frequencies 
are almost unaffected. In addition, we observed a power-law correction with an interaction-dependent exponent. Remarkably, our functional RG procedure thus (automatically and consistently) carries out two different types of resummations: one leads to the combined appearance of the time $t$ 
and interaction $U$ in the argument of exponential functions; the other gives rise to a 
power-law in $t$ with a $U$-dependent exponent. This is highly nontrivial. We compared our results to those obtained within a recently-developed approximate real-time RG approach and observed good agreement. 
  
We then exploited a mapping of the unbiased IRLM to the ohmic spin-boson model. The latter features two localized states; if their energy is equal, one expects a coherent-decoherent transition, which is confirmed by our calculation. The FRG prediction for the relation between the relaxation rate and the oscillation frequency agrees with those obtained by a field theoretical approach and an improved noninteracting blib approximation. Note that this relation is a manifestation of strong-coupling physics: The explicit scale set by the energy difference of the two levels is zero, and the coherent-decoherent transition is purely driven by the interaction (the oscillation frequency is proportional to $U$). Reproducing these results hence provides another stringent test for our newly-developed method.   

The time-dependent functional RG approach directly allows to tackle explicitly time-dependent Hamiltonians. This constitutes a highly-active field of current research, and one can readily envisage a vast number of applications. From a theoretical perspective, it is intriguing to investigate the time evolution of systems with {\it correlated initial density matrices} towards their steady-state through the following protocol: one starts with an uncorrelated initial density matrix; the system then relaxes towards a steady-state which contains correlations; thereafter, the system is quenched, and one studies the relaxation process out of the now correlated initial state. From a practical point of view, one can investigate correlation effects on quantum pumps (which require a periodic variation of the dot parameters).

\section*{Acknowledgement}
We thank S.\ Andergassen, M.\ Pletyukhov, H.\ Schoeller, and D.\ Schuricht for very useful discussions. We are grateful to D.\ Schuricht for providing the RTRG data of Figs.\ \ref{fig:RTRGvsFRGocc} and \ref{fig:RTRGvsFRGcur}. This work was supported by the DFG via FOR 723 and KA3360-1/1 (C.K.).  

\appendix*
\setcounter{equation}{0}
\section*{Appendix}
\label{appen}

To solve the recursion Eq.\ \eqref{eq:GKdis} we need to compute the integral
\begin{equation}
\int \limits_{t_n}^{t_{n+1}}ds_1\int\limits_{t_m}^{t_{m+1}}ds_2 G^{\Ret}(t_{n+1},s_1)\Sigma^\K_\res(s_1,s_2) G^{\Adv}(s_2,t_{m+1}) 
\label{intappendix}
 \end{equation}
with $\Sigma^{\K}_\res(s_1,s_2)$ as in Eq.\ \eqref{eq:SigmaKinter}. 
The exponential of the matrix in Eq.\  \eqref{eq:GRetdisc} can be evaluated:
\begin{equation}
G^{\Ret}_{ij}(t,t')=-i \sum\limits_{l=1}^3\text{Res}_{ij,l}e^{-i\omega_l(t-t')},
\end{equation}
with $ \text{Res}_{ij,l} $ and $ \omega_l $ being the residues and the poles of 
\begin{equation}
\label{eq:forresiduesandpoles}
\frac{1}{\omega-\left(\tilde h_0^{\rm dot} +\tilde \Sigma^\Ret_{\res}-i\Lambda+\tilde \Sigma^\Ret_{\bar t}\right)}
\end{equation}
Introducing the effective parameters
\begin{equation}
\begin{split}
\epsilon^{\prime\Lambda}& = \Sigma^\Ret_{\bar t,11}-U/2 , \\
\epsilon^{\Lambda}& = \epsilon+\Sigma^\Ret_{\bar t,22}-U , \\
\tau_{12}^\Lambda& = \tau+\Sigma^\Ret_{\bar t,12} , \\
\tau_{23}^\Lambda& = \tau+\Sigma^\Ret_{\bar t,23}  
\end{split}
\end{equation}
allows to express the poles as
\begin{widetext}
\begin{align}
\omega_1 =\epsilon^{\prime\Lambda}-i(\Gamma+\Lambda) \; , \;\;\; 
\omega_{2/3} = \frac{1}{2}\left(\epsilon^\Lambda+ \epsilon^{\prime\Lambda}-i\Gamma-2i\Lambda\mp 
\sqrt{-(\Gamma-i\epsilon^\Lambda+i\epsilon^{\prime,\Lambda})^2+4|\tau_{12}^\Lambda|^2+4|\tau_{23}^\Lambda|^2}\right)~.
\end{align} 
The corresponding residues are given in Tab.\ \ref{tab:residues}.

\begin{table}
\centering
\begin{tabular}{ccccccccc}
\multicolumn{2}{c}{$\text{Res}_{ij,n}$}&\multicolumn{4}{c}{}&{\color{white}space}&$ij$&\\\noalign{\vskip0.1cm}
\multicolumn{2}{c}{}&\multicolumn{4}{c}{$ij$}&{\color{white}space}&33&{$\text{Res}_{11,n}(\tau_{12}^\Lambda\to \tau_{23}^\Lambda)$}\\\noalign{\vskip0.1cm}
\cline{8-9}
\noalign{\vskip0.1cm}
&&11&22&12&13&&21&{$\text{Res}_{12,n}(\tau_{12}^\Lambda\to (\tau_{12}^\Lambda)^*)$}\\
\noalign{\vskip0.1cm}
\cline{1-6}\cline{8-9}
\noalign{\vskip0.1cm}
&1&$1+\frac{|\tau_{12}^\Lambda|^2}{(\omega_1-\omega_2)(\omega_1-\omega_3)}$&$0$&$0$&$\frac{\tau_{12}^\Lambda \tau_{23}^\Lambda}{(\omega_1-\omega_2)(\omega_1-\omega_3)}$&&31&{$\text{Res}_{13,n}(\tau_{12}^\Lambda,\tau_{23}^\Lambda\to (\tau_{12}^\Lambda)^*,(\tau_{23}^\Lambda)^*)$}\\\noalign{\vskip0.1cm}
\cline{8-9}
\noalign{\vskip0.1cm}
n&2&$\frac{|t_{12}^\Lambda|^2}{(\omega_2-\omega_1)(\omega_2-\omega_3)}$&$\frac{\omega_2-\omega_1}{\omega_2-\omega_3}$&$\frac{\tau_{12}^\Lambda}{\omega_2-\omega_3}$&$\frac{\tau_{12}^\Lambda \tau_{23}^\Lambda}{(\omega_2-\omega_1)(\omega_2-\omega_3)}$&&23&{$\text{Res}_{12,n}(\tau_{12}^\Lambda\to \tau_{23}^\Lambda)$}\\\noalign{\vskip0.1cm}
\cline{8-9}
\noalign{\vskip0.1cm}
&3&$\frac{|\tau_{12}^\Lambda|^2}{(\omega_3-\omega_1)(\omega_3-\omega_2)}$&$\frac{\omega_3-\omega_1}{\omega_3-\omega_2}$&$\frac{\tau_{12}^\Lambda}{\omega_3-\omega_2}$&$\frac{\tau_{12}^\Lambda \tau_{23}^\Lambda}{(\omega_3-\omega_1)(\omega_3-\omega_2)}$&&32&{$\text{Res}_{23,n}(\tau_{23}^\Lambda\to (\tau_{23}^\Lambda)^*)$}\\

\end{tabular}\\ 
\caption{Residues of Eq.\ \eqref{eq:forresiduesandpoles}.}
\label{tab:residues}
\end{table}

To compute the integral Eq.\ (\ref{intappendix}) one substitutes $T =t_1+t_2$ and $\Delta t =t_2-t_1$:
\begin{equation}
\int \limits_{t_n}^{t_{n+1}}dt_1\int\limits_{m_j}^{t_{m+1}}dt_2\longrightarrow\frac{1}{2}\left[\;\int\limits_{t_m-t_{n+1}}^{t_{m+1}-t_{n+1}}d\Delta t\int\limits_{2t_m-\Delta t}^{2t_{n+1}+\Delta t}dT +\int\limits_{t_{m+1}-t_{n+1}}^{t_{m}-t_n}d\Delta t\int\limits_{2t_m-\Delta t}^{2t_{m+1}-\Delta t}dT +\int\limits_{t_m-t_{n}}^{t_{m+1}-t_n}d\Delta t\int\limits_{2t_n+\Delta t}^{2t_{m+1}-\Delta t}dT \right]
\end{equation}
for $ t_{n+1}-t_{n}\geq t_{m+1}-t_m $ (the opposite case follows analogously). Because of the principal value involved it proves advantageous to separate the 
problem into the two cases $  n=m $ and $  n\neq m $. The case $n=m$ also includes the solution of the noninteracting system, where one can choose a single discretization step; we will illustrate it first.
With the above substitution one can write the $i,j$ matrix element of Eq.\ (\ref{intappendix}) in terms of exponential integrals:\cite{abramowitz}
\begin{equation}
\begin{split}
&\left[ \int \limits_{t_n}^{t_{n+1}}ds_1\int\limits_{t_n}^{t_{n+1}}ds_2 G^{\Ret}(t_{n+1},s_1)\Sigma^\K_\res(s_1,s_2) 
G^{\Adv}(s_2,t_{n+1}) \right]_{ij}
=\lim\limits_{\delta\to 0}\frac{1}{\pi}\sum\limits_{\alpha=L,R\atop n,m=1,2,3}
\text{Res}_{i \alpha,n} \text{Res}_{j \alpha,m}^* e^{-i\Delta \omega_{nm}t_{n+1}}\\
&\times\left[\;\int\limits_{t_n-t_{n+1}}^{-\delta}d\Delta t\int\limits_{2t_n-\Delta t}^{2t_{n+1}+\Delta t}dT +\int\limits_{\delta}^{t_{n+1}-t_n}d\Delta t\int\limits_{2t_n+\Delta t}^{2t_{n+1}-\Delta t}dT \right] \frac{\Gamma}{\Delta t } e^{i\mu_\alpha\Delta t } e^{\frac{1}{2}i T \Delta \omega_{nm}} e^{-i\frac{1}{2}\Delta t (\omega_n+\omega_m^*)}\\
&=\sum\limits_{{\alpha=L,R} \atop{n,m}=1,2,3}\text{Res}_{i \alpha, n} \text{Res}^*_{ j \alpha,m}e^{-i(\omega_{n}t_{n+1}-\omega_m^*t_{n+1})}\frac{2\Gamma}{i\Delta \omega_{nm}\pi}\\
&\phantom{=}\times\bigg[e^{i\Delta\omega_{nm}t_{n+1}}\left\{-\log(-[\mu_\alpha-\omega_n])+\log(\mu_\alpha-\omega_m^*)\right\}+e^{i\Delta\omega_{nm}t_{n+1}}E_1(-i[\mu_\alpha-\omega_m^*][t_n-t_{n+1}])\\
&\phantom{=}\phantom{\times\bigg[}-e^{i\Delta\omega_{nm}t_{n}}\left\{-\log(-[\mu_\alpha-\omega_m^*])+\log(\mu_\alpha-\omega_n)\right\}-e^{i\Delta\omega_{nm}t_{n}}E_1(-i[\mu_\alpha-\omega_n][t_n-t_{n+1}])\\
&\phantom{=}\phantom{\times\bigg[}-e^{i\Delta\omega_{nm}t_{n+1}}E_1(-i[\mu_\alpha-\omega_n][t_{n+1}-t_n])
+e^{i\Delta\omega_{nm}t_{n}}E_1(-i[\mu_\alpha-\omega_m^*][t_{n+1}-t_n])\bigg],\label{eq:GK3sites1}
\end{split}
\end{equation}
where one has exploited that
\begin{equation}
\lim\limits_{\delta\to0^+}E_1(-x\delta)-E_1(y\delta)=-\log(-x)+\log(y)\label{eq:E1limit}~,
\end{equation}
and defined
\begin{equation}
\Delta\omega_{nm} =  \omega_n-\omega_m^*\label{eq:Deltaw}.
\end{equation}
In the indices of the residues in Eq.\ (\ref{eq:GK3sites1}) one has to replace $\alpha=L$ by $1$ and $\alpha=R$ by $3$. 
For the case $m\neq n$ one analogously finds
\begin{equation}
\begin{split}
&\left[ \int \limits_{t_n}^{t_{n+1}}ds_1\int\limits_{t_m}^{t_{m+1}}ds_2 G^{\Ret}(t_{n+1},s_1)\Sigma^\K_\res(s_1,s_2) 
G^{\Adv}(s_2,t_{m+1}) \right]_{ij}
=\frac{1}{\pi}\sum\limits_{\alpha=L,R\atop n,m=1,2,3}\text{Res}_{i\alpha,n}\text{Res}_{j\alpha,m}^* 
e^{-i(\omega_{n}t_{n+1}-\omega_m^*t_{m+1})}\\
&\times\left[\;\int\limits_{t_m-t_{n+1}}^{t_{m+1}-t_{n+1}}d\Delta t\int\limits_{2t_m-\Delta t}^{2t_{n+1}+\Delta t}dT +\int\limits_{t_{m+1}-t_{n+1}}^{t_{m}-t_n}d\Delta t\int\limits_{2t_m-\Delta t}^{2t_{m+1}-\Delta t}dT +\int\limits_{t_m-t_{n}}^{t_{m+1}-t_n}d\Delta t\int\limits_{2t_n+\Delta t}^{2t_{m+1}-\Delta t}dT \right] 
\frac{\Gamma}{\Delta t } e^{i\mu_\alpha\Delta t } e^{\frac{1}{2}i T \Delta \omega_{nm}} 
e^{-i\frac{1}{2}\Delta t (\omega_n+\omega_m^*)}\\
&=\sum\limits_{{\alpha=L,R}\atop{n,m}=1,2,3}\text{Res}_{i\alpha,n} \text{Res}^*_{ j\alpha,m}e^{-i(\omega_{n}t_{n+1}-\omega_m^*t_{m+1})}\frac{2\Gamma}{i\Delta \omega_{nm}\pi}\\
&\phantom{=}\times\bigg[-e^{i\Delta\omega_{nm}t_{n+1}}E_1(-i[\mu_\alpha-\omega_m^*][t_{m+1}-t_{n+1}])+e^{i\Delta\omega_{nm}t_{n+1}}E_1(-i[\mu_\alpha-\omega_m^*][t_m-t_{n+1}])\\
&\phantom{=}\phantom{\times\bigg[}+e^{i\Delta\omega_{nm}t_{m}}E_1(-i[\mu_\alpha-\omega_n][t_m-t_n])-e^{i\Delta\omega_{nm}t_{m}}E_1(-i[\mu_\alpha-\omega_n][t_m-t_{n+1}])\\
&\phantom{=}\phantom{\times\bigg[}-e^{i\Delta\omega_{nm}t_{m+1}}E_1(-i[\mu_\alpha-\omega_n][t_{m+1}-t_n])+e^{i\Delta\omega_{nm}t_{m+1}}E_1(-i[\mu_\alpha-\omega_n][t_{m+1}-t_{n+1}])\\
&\phantom{=}\phantom{\times\bigg[}+e^{i\Delta\omega_{nm}t_{n}}E_1(-i[\mu_\alpha-\omega_m^*][t_{m+1}-t_n])-e^{i\Delta\omega_{nm}t_{n}}E_1(-i[\mu_\alpha-\omega_m^*][t_m-t_{n}])\bigg].\label{eq:GK3sites1_a}
\end{split}
\end{equation}
\end{widetext}




\begin{thebibliography}{99}
\bibitem{commentnumpara}
The results presented in in Sec.\ \ref{sec:resultsint} were calculated using the following discretization parameters: The total number $M$ of 
discretization points is $501$, $ n=0\dots 500 $ (leading to a system of the order of $10^3$ coupled differential equations). Of those $11$ constitute the first part of the mesh $t_{n\leq 10}=2n/(10\Gamma)$ to account for the small time scale $ \sim 1/\Gamma $. The remaining $ 490 $ are used to incorporate the larger scale $ t_{11\leq n\leq 500}=8(n-10)/\Gamma+2/\Gamma$.
\bibitem{fehler} Note that Eq.\ (53) of Ref.\ \onlinecite{Anders2} contains a typo. 
The prefactor 2 in the first line of Eq.\ (53) must be erased.
\end{thebibliography}
\end{document}